\documentclass[aps,nofootinbib,floatfix]{revtex4}
\newif\ifpdf
	\ifx\pdfoutput\undefined
	\pdffalse 
	\else
	\pdfoutput=1 
	\pdftrue
\fi

\ifpdf
	\usepackage[pdftex]{graphicx}
	\else
	\usepackage{graphicx}
\fi
\newcommand{\be}{\begin{equation}}
\newcommand{\ee}{\end{equation}}
\newcommand{\bea}{\begin{eqnarray}}
\newcommand{\eea}{\end{eqnarray}}
\newcommand{\nn}{\nonumber}

\newcommand{\D}{\displaystyle}

\newcommand{\ha}{\hat{a}}

\newcommand{\had}{\hat{a}^\dagger}

\newcommand{\bep}{{\bf e}}
\newcommand{\bepsa}{{\bf e}^{(a)}}

\newcommand{\ci}{{\rm i}}
\newcommand{\bdh}{{\hat{\bf d}}}
\newcommand{\brh}{{\hat{\bf r}}}
\newcommand{\bE}{{\bf E}}
\newcommand {\Deltabar}{\overline{\Delta}}

\newcommand{\ket}[1]{{|#1\rangle}}
\newcommand{\bra}[1]{{\langle#1|}}

\newcommand{\om}[1]{{{\omega_{#1}}}}
\newcommand{\threej}[6]{\left(\begin{array}
{ccc}#1&#2&#3\\#4&#5&#6\end{array}\right)}
\begin{document}
\ifpdf
	\DeclareGraphicsExtensions{.pdf, .jpg, .tif}
	\else
	\DeclareGraphicsExtensions{.eps, .jpg}
\fi
\title{A Raman approach to quantum logic in Calcium-like ions}
\author{Mark S. Gulley}
\affiliation{LANSCE-6, Los Alamos National Laboratory, Los Alamos, NM 87545, USA}
\author{Andrew G. White}\email{http://www.physics.uq.edu.au/qt, andrew@physics.uq.edu.au}
\affiliation{Department of Physics, University of Queensland, Brisbane, Queensland 4072, AUSTRALIA}
\author{Daniel F. V. James}
\affiliation{T-4, Theoretical Division,Los Alamos National Laboratory, Los Alamos, NM 87545, USA}

\begin{abstract}
We consider the feasibility of performing quantum logic operations based on stimulated Raman transitions in trapped Calcium ions. This technique avoids many of the technical difficulties involved with laser stabilisation, and only three laser wavelengths are required, none of which need have particularly stringent requirements on their bandwidths. The possible problems with experimental realisations are discussed in detail.
\end{abstract}
\maketitle

\section{Introduction}

Quantum information processing is a new paradigm in computation, and the possibility of realizing it via trapped ions is currently a subject of active research in a number of groups worldwide.  The basic idea of trapped ion quantum computation is to use two internal quantum states of a trapped ion to form a ``qubit'' for the storage of information, and to employ the strongly coupled quantum oscillations of the ions to perform the conditional dynamics ( or ``quantum logic gates'') required to execute quantum algorithms~\cite{CZ,NISTgate}.  The quantum dynamics of both the internal and external degrees of freedom of the ions are controlled by pulsed laser beams.  Recent reviews of the experimental issues involved are given in refs.~\cite{thepaper, NISTreview}.

There are two different ways of performing the laser-induced transitions required for quantum computation.  Conceptually the most simple is to use the Rabi oscillations between two internal states of the ion.  If the ion interacts with a monochromatic field (i.e. a single narrow band laser beam) precisely tuned to the transition frequency for the two levels, the population will oscillate back and forth between them~\cite {AllenEberly}.  A long-lived two level system is required to make a qubit suitable for quantum computation, and so atomic levels with dipole-forbidden transitions are the most suitable. An example of such a dipole forbidden transition is the 729 nm $4\,^{2}{\rm S}_{1/2}$ to $3\,^{2}{\rm D}_{5/2}$ transition in ${\rm Ca}^{+}$ (fig.1.1), which has a natural lifetime of about 1.1 seconds~\cite{DFVJ}.  We will refer to this method as the ``single laser'' technique.  The theoretical background is described in some detail in ref.~\cite{DFVJ}.

Alternatively, qubits can be manipulated using stimulated Raman transitions.  The advantages of using stimulated Raman transitions, instead of single-photon transitions, are now well known.  Two lasers, traditionally named the ``pump'' and the ``Stokes'' beams, are tuned so that population from one level is excited to some intermediate virtual level by one laser, and then immediately brought back down from the virtual level to a different atomic state by the second laser.  Following the first demonstration of these transitions between atomic ground states~\cite{Thomas}, Raman transitions were proposed for trapped ion experiments~\cite{Heinzen}, then used to cool trapped 
neutral atoms~\cite{Kasevich}.  Raman transitions have been used in ion traps in a series of experiments by the NIST group \cite{NISTraman}.  In this paper we discuss in detail the use of such stimulated Raman transitions to perform quantum logic operations in ion-trap quantum computers based on ${\rm Ca}^{+}$, or ${\rm Ca}^{+}$-like ions.
In particular, we focus on two issues: options for realising quantum logic operations; and measurement of the final state, i.e. the ``read out''.  The latter issue has been investigated in $^{40} \rm{Ca}^{+}$ \cite{Stevens:98}, however that method does not realise an an ideal von Neumann measurement (projection onto an orthonormal basis).

The principal advantages of the Raman scheme in comparison with the single laser scheme are the much reduced optical stability requirements and a reduction in the rate of spontaneous emission \cite{Shore:90}.  Only high {\it relative} frequency stability of the pump and Stokes beams is required, which is easily achieved with current technologies.  The two beams can be generated by dividing a parent beam into two with a beam-splitter and modulating one of the resultant beams using an electro-optic modulator (EOM) driven by a stable radio-frequency (RF) source.  Regarding spontaneous emission, both the lower and the upper qubit states $\ket{0}$ and $\ket{1}$ are sublevels of the ground state, and so the radiative lifetime of $\ket{1}$ is extremely large ($\sim 10^{6}$ years or more).  As all logical operations must be carried out before spontaneous emission of the $\ket{1}$ state can occur, this significantly relaxes the operational constraints.  The error rate per CNOT operation has been calculated to be two to three orders of magnitude smaller in Raman schemes than in single laser transition schemes~\cite{thepaper}.

The choice of ion species is driven principally by the availability of reliable lasers at the requisite wavelengths.  Figure 1.1 shows the level scheme for $^{40}{\rm Ca}^{+}$, when the $4\,^{2}{\rm S}_{1/2}$ ground state level is Zeeman-split into sublevels by a modest magnetic field (the splitting is $2\pi \times 2.8$MHz per Gauss~\cite{Woodgate}).  The ground state sublevels of each ion are assigned to be the two levels of each qubit (i.e. the logical $\ket{0}$ and $\ket{1}$ states).  The two Raman lasers would then be used to perform transitions between these two states, via a virtual level.  For singly ionised calcium (${\rm Ca}^{+}$) the basic wavelength ranges required are 393-397~nm, 729-732~nm and 849-866~nm, which can be realised relatively easily with existing commercially available systems, and (for the 393-397~nm range) a doubling system. Although ${\rm Ca}^{+}$ is attractive in terms of wavelength, the most abundant isotopes of Ca$^{+}$ have zero nuclear spin and thus no hyperfine splitting.  This is unfortunate because the hyperfine structure of the ground state of other species (e.g. $^{9}{\rm Be}^{+}$, $^{137}{\rm Ba}^{+}$, $^{25}{\rm Mg}^{+}$ ) provides both a large number of additional quantum states and the frequency splitting required to perform quantum logic efficiently using Raman transitions. However, as discussed in section 3, there are a number of options available for by-passing this problem.

The paper is organised as follows.  In section 2 we present a concise review of the theoretical description of stimulated Raman transitions as applied to trapped atoms or ions, for which both internal and external degrees of freedom must be considered.  Section 3 explains how Raman transitions can be used to perform all of the tasks required for quantum computation.  Section 4 discusses the experimental issues we anticipate in carrying out this program in the laboratory, and section 5 states our conclusions.

\section{Theory of Stimulated Raman Transitions}
\setcounter{equation}{0}

\subsection{Basic notions and notation}
Theoretical treatment of stimulated Raman processes in atoms has been considered by many authors (see, for example, the comprehensive works by Schenzle and Brewer~\cite{Schenzle:77} and by Shore~\cite{Shore:90}).  For the convenience of the reader, we recapitulate the relevant theory here.

The quantum states of a trapped ion consist of a tensor product of the wavefunction describing the internal states of the ion and of the wavefunction describing the external vibrational state of the ion in the trap.  The internal states are specified by a set of quantum numbers; we will denote these states by the short-hand notation $\ket{\gamma, J, m_J}$, where ${J, m_J}$ are the usual angular momentum quantum numbers and $\gamma$ lumps together all of the other quantum numbers which specify the state.  As discussed below, Calcium has only one very rare stable isotope with non-zero nuclear spin. Thus we will assume that we are dealing with spin zero nuclei, so there is no hyperfine structure to consider, and ${J m_J}$ are ``good'' quantum numbers.  We shall divide these internal states of the ion into ``upper'' and ``lower'' manifolds.  What we have in mind for these manifolds are the various magnetic sublevels of some atomic level: For example, the lower manifold could be the two sub-levels of the $4\,^{2}{\rm S}_{1/2}$ ground state, while the upper manifold could be the two sublevels of the $4\,^{2}{\rm P}_{1/2}$ or the $4\,^{2}{\rm P}_{3/2}$ excited levels, although we shall develop the theory for arbitrary upper and lower manifolds.  Further, we will assume that transitions between any state in the upper manifold and any state in the lower manifold are dipole allowed, and that transitions between any pair of states in the upper manifold or any pair of states in the lower manifold are forbidden.  Generalizing the theory to consider the more exotic case when transitions between the upper and lower manifolds are dipole forbidden (e.g. when the upper manifold is the is $3\,^{2}{\rm D}_{3/2}$ level) is reasonably easy. This situation may have eventual application to ion trap quantum computers, because the performance of such devices will be limited by spontaneous emission from the upper manifold, as has been pointed out by Plenio and Knight~\cite{PK}; thus if a long-lived upper manifold can be used, in principle higher performance could be obtained.

We will assume that the ions are confined in a highly anisotropic trap, with very strong confining potentials in two directions and a comparatively weak potential in the third, axial direction, which we will refer to as the x-axis.  Provided that the the trapping potentials have a sufficiently high degree of anisotropy~\cite{schiffer}, and the number of ions in the trap is not too large, the ions will lie in a string along the x-axis.  The vibrational motion of the different ions will be strongly coupled due to the Coulomb repulsion, and so their motion, when small, is most easily described in terms of a set of normal modes.  The mode angular frequencies are given by $\sqrt{\mu_{p}}\omega_{x}$, where $\mu_{p}$ is the eigenvalue associated with the mode~\cite{DFVJ}, and $\omega_{x}$ is the angular frequency characterizing the strength of the trapping potential in the x direction.  For simplicity, we will only consider oscillation along the x direction, oscillations transverse to this direction are important only for low anisotropies if necessary the theory presented here can be extended to this case without too much difficulty).  The vibrational quantum state of the ions is specified by a set of excitation numbers for the various phonon modes of the oscillation.  If there is a total of $N$ ions in the trap, then there will be $N$ such modes, which can be enumerated in order of increasing eigenfrequency (see ref.~\cite{DFVJ} for more details).  Each of these modes can be considered as an uncoupled quantum harmonic oscillator, and so each has a ladder of number states.  Thus the state will be denoted $\ket{\{n\}}$, where $\{n\}n_{1}, n_{2},\ldots n_{N}$ is the set of mode excitation numbers, i.e. there are $n_{1}$ quanta in the first mode, $n_{2}$ quanta in the second mode, and so on.

We will use a single index to denote the combined wavefunction for the internal state and for the vibrational state. Thus $\ket{i}$ will stand for $\ket{\gamma, J, m_J}\otimes \ket{n_{1}, n_{2},\ldots n_{N}}$.  We shall use $\{\ket{i},\ket{j}, \ket{i'},\ket{j'}\; \mbox{etc.}\}$ to denote states in the lower manifold of atomic states and $\{\ket{k},\ket{l}, \ket{k'},\ket{l'}\; \mbox{etc.}\}$ to denote states in the upper manifold of atomic states.  The energies of these states will be denoted $ \hbar \om{i}, \hbar \om{k}$ etc.  Thus the state function for the ion will be given in the Schr{\"o}dinger picture by the following expression \be \ket{\psi(t)}=\sum_{j}a_{j}(t) \exp\left(-\ci\om{j}t\right)\ket{j} +\sum_{k}b_{k}(t) \exp\left(-\ci\om{k}t\right)\ket{k}\label{psi1} , \ee where $a_{j}$ is the amplitude in a lower manifold state $\ket{j}$, and $b_{k}$ is the amplitude in an upper manifold state $\ket{k}$, the summations being over all lower manifold states and all upper manifold states, respectively.  We will make the `essential states' approximation, i.e. assume that these states are the only ones which will have a non-zero amplitude; thus the following equations apply:
\bea
\hat{I}&=&\left\{\sum_{j}\ket{j}\bra{j}+\sum_{k}\ket{k}\bra{k}\right\}
\label{ident} \\
\hat{H}_{0}&=&\hbar\left\{\sum_{j}\om{j}\ket{j}\bra{j}+
\sum_{k}\om{k}\ket{k}\bra{k}\right\} ,
\eea
where $\hat{I}$ is the identity operator and $\hat{H}_{0}$ is the unperturbed Hamiltonian operator.

The interaction Hamiltonian describing the interaction of the atomic system with the laser field is given by
\be
\hat{H}_{I}=-\bdh\cdot \bE\left(\brh,t\right)
\label{Hi1},
\ee
where $\hat{\bf{d}}$ is the dipole moment operator for the ion (which acts only on the internal atomic states), $\bf{E}$ is the electric field strength, which is a function of time $t$ and of the position of the ion, which is expressed in terms of its operator $\hat{\bf{r}}$, which acts on the vibrational states. The electric field will be assumed to consist of a sum of monochromatic components,
\be
\bE \left(\brh,t\right) =  Re\left\{\sum_{a}
\bepsa E^{(a)}(\brh) \exp(\ci\om{a}t)\right\},
\label{Eee}
\ee
where $\bepsa$ is the unit polarisation vector for the $a-th$ laser beam. Substituting eq.~(\ref{Eee}) into eq.~(\ref{Hi1}), and using (\ref{ident}) and the fact that, by assumption $\bra{k}\hat{\bf{d}}\ket{k'}=\bra{j}\hat{\bf{d}}\ket{j'}=0$, and making the rotating wave approximation (i.e. neglecting terms oscillating at frequencies $\om{k}-\om{j}+\om{a}$) one obtains the following formula for the interaction Hamiltonian:
\be
\hat{H}_{I}=-\sum_{jk}\sum_{a}\frac{\hbar}{2}
\Omega^{(a)}_{jk}
\ket{j}\bra{k}\exp(\ci\om{a}t)
+h.a. , \label{Hi2}
\ee
where
\be
\hbar\Omega^{(a)}_{jk}=
\bra{j}\hat{\bf{d}}\cdot\bepsa
E^{(a)}(\hat{\bf{r}}) \ket{k} .
\label{parker}
\ee
Thus one obtains the following differential equations for the amplitudes $a_{j}(t)$ and $b_{k}(t)$:
\bea
\dot{a}_{j}(t)&=&\frac{\ci}{2}\sum_{k}\sum_{a}
\Omega^{(a)}_{jk}
\exp\left(-\ci\Delta^{(a)}_{jk}t\right)
b_{k}(t) ,\label{adot}\\
\dot{b}_{k}(t)&=&\frac{\ci}{2}\sum_{j'}\sum_{a}
\Omega^{(a)\ast}_{j'k}
\exp\left(\ci\Delta^{(a)}_{j'k}t\right)
a_{j'}(t) \label{bdot} ,
\eea
where the asterisk denotes complex conjugation and
\be
\Delta^{(a)}_{jk}=\om{k}-\om{j}-\om{a} ,
\ee
(see fig.2.1).

If we adiabatically eliminate the populations in the upper manifold (see Appendix A), one obtains the following formula for the amplitudes in the lower manifold:
\be
\dot{a}_{i}(t)=\ci\sum_{j}\beta_{ij}(t)a_{j}(t) ,
\label{adot2}
\ee
where the coupling matrix $\beta_{ij}(t)$ is given by
the following formula:
\be
\beta_{ij}(t)=\sum_{k}\sum_{a,a'}
\frac{\Omega^{(a)}_{ik}\Omega^{(a')\ast}_{jk}}
{4\Delta^{(a')}_{jk}}
\exp\left[\ci\left(\om{ij}-\om{a'a}\right)t\right] .
\label{beta}
\ee

\subsection{Raman transitions between two levels with two laser fields}
We will now examine the special case in which there are only {\em two} laser beams present, the ``pump'' field and the ``Stokes'' field, with frequencies $\om{p}$ and $\om{s}$ respectively, such that $\om{ps}$$\equiv$$\om{p}-\om{s}$$>0$.  Thus using eq.~(\ref{beta}), and neglecting ``rapidly'' oscillating terms, (i.e those proportional to $\exp\left[\pm \ci\left(\om{ps}\right)t\right]$), one obtains the following expressions for the elements of the matrix
$\beta_{ij}(t)$:
\bea
\beta_{ii}&=&\sum_{k}
\frac{\left|\Omega^{(p)}_{ik}\right|^{2}}{4\Delta^{(p)}_{ik}}+
\frac{\left|\Omega^{(s)}_{ik}\right|^{2}}{4\Delta^{(s)}_{ik}}, \nn \\
&& \nn \\
\beta_{ij}(t)&=&\left\{
\begin{array}{ll}
\displaystyle
\left[\sum_{k}\frac{\Omega^{(p)}_{ik}\Omega^{(s)\ast}_{jk}}
{4\Delta^{(s)}_{jk}}\right] \exp\left[\ci
t(\om{ps}-|\om{ij}|)\right]& (\om{j}>\om{i}),\\
&\\
\displaystyle
\left[\sum_{k}
\frac
{\Omega^{(p)\ast}_{jk}\Omega^{(s)}_{ik}}
{4\Delta^{(p)}_{jk}}
\right]
\exp\left[-\ci t(\om{ps}-|\om{ij}|)\right]& (\om{i}>\om{j}).
\end{array}\right. \nn \\
\eea
The constant diagonal elements represent an ``A.C. Stark'' shift, i.e. a change in the resonance frequency of a transition due to an applied optical field.  The off-diagonal elements represent induced transitions between pairs of levels of the lower manifold.

We will now make a further specialisation by considering only two lower manifold states.  We will assume that the quantity $\om{ps}-|\om{ij}|$ is large for all but a single pair of states which we will denote $\ket{I}$ and $\ket{J}$, where $(\om{J}>\om{I})$.  Thus the off-diagonal elements of the matrix $\beta_{ij}$ will be very rapidly oscillating except for these two states, and there will be negligible probability of any transitions being induced.  We will denote the average detuning of the $k$-th upper manifold state by $\Deltabar_{k}=\om{k}-(\om{I} +\om{J} + \om{s} + \om{p})/2$, and the Raman detuning by $\delta = \om{ps}-\om{JI}$ (see fig.2.2). If one makes the assumption that $\Deltabar_{k}\gg\delta$ and $\Deltabar_{k}\gg\om{ps}$, then the elements of the $2\times 2$ matrix $\beta_{IJ}$ can be written as:
\bea
\beta_{II}&\approx&\sum_{k}
\frac{\left|\Omega^{(p)}_{Ik}\right|^{2}+\left|\Omega^{(s)}_{Ik}\right|^{2}}
{4\Deltabar_{k}}\equiv A \\
\beta_{JJ}&\approx&\sum_{k}
\frac{\left|\Omega^{(p)}_{Jk}\right|^{2}+\left|\Omega^{(s)}_{Jk}\right|^{2}}
{4\Deltabar_{k}}\equiv D \\
\beta_{IJ}(t)&\approx&\beta_{JI}(t)^{\ast}\approx\left[\sum_{k}
\frac{\Omega^{(p)}_{Ik}\Omega^{(s)\ast}_{Jk}}
{4\Deltabar_{k}}\right]\exp\left(\ci t\delta \right)
\equiv B \exp\left( \ci t\delta \right).
\eea
The amplitudes for the two states $\ket{I}$ and $\ket{J}$ therefore obey the following equation
\be
\frac{\partial}{\partial t}
\left(\begin{array}{c}
a_{I}\\
a_{J}\end{array}\right) =
\ci \left(
\begin{array}{cc}
A & B e^{\ci t \delta}\\
B^{\ast} e^{-\ci t \delta} & D \end{array}
\right)
\left(\begin{array}{c}
a_{I}\\
a_{J}\end{array}\right).
\label{haigh}
\ee
These are equivalent to the well-known Bloch equations for a two level system~\cite{AllenEberly,FVH}. If one makes the following resonance condition
\be
\delta = A-D,
\ee
which can be achieved in practice by varying the relative frequency of the two laser beams, then the solution of equation (\ref {haigh}) is equivalent to the standard on-resonance solution of the Bloch equations.  Using this solution we can obtain the following solutions for the amplitudes
$a_{I}(t)$ and $a_{J}(t)$:
\be
\left(\begin{array}{c}
a_{I}(t)\\
a_{J}(t)\end{array}\right) =
\exp\left[\ci (A+D)t/2 \right]
\left(
\begin{array}{cc}
e^{\ci \chi}\cos\theta/2 & \ci e^{\ci \phi} \sin\theta/2\\
\ci e^{-\ci \phi} \sin\theta/2 &e^{-\ci \chi}\cos\theta/2 \end{array}
\right)
\left(\begin{array}{c}
a_{I}(0)\\
a_{J}(0)\end{array}\right),
\ee
where
\bea
\theta&=&2\left|\Omega_{eff}\right|t\label{thetadef} \\
\phi&=& \arg\left\{\Omega_{eff}\right\}+\chi \\
\chi&=&\frac{t\delta}{2}=
\sum_{k}
\frac{\left(
\left|\Omega^{(p)}_{Ik}\right|^{2}+
\left|\Omega^{(s)}_{Ik}\right|^{2}-
\left|\Omega^{(p)}_{Jk}\right|^{2}-
\left|\Omega^{(s)}_{Jk}\right|^{2}\right)t}
{8\Deltabar_{k}},
\eea
and the effective Rabi frequency for Raman transitions between levels $\ket{I}$ and $\ket{J}$ is given by the formula
\be
\Omega_{eff}=\sum_{k}
\frac{\Omega^{(p)}_{Ik}\Omega^{(s)\ast}_{Jk}}
{4\Deltabar_{k}}.
\label{hulme}
\ee
Thus by controlling the duration of each laser pulse precisely, one can control the populations in the two levels $\ket{I}$ and $\ket{J}$.  Furthermore, by controlling the relative phases of the pump and Stokes lasers one can control the relative phases of the two states.  The phase change $\chi$, caused by the A.C. Stark shift, will be small, but must be taken into account if large scale quantum computations are ever to be attempted in ion traps.

The value of $\Omega_{eff}$ can be calculated in terms of atomic constants as shown in Appendix B. In quantum computation the phonon modes are assumed to be unpopulated.  Two types of transition are of particular use, namely the ``V'' type or carrier transition, for which phonon state is left unchanged and the ``U'' type or red sideband transition, in which one phonon is created \cite{CZ}.  To be more specific, for a ``V'' pulse,
\bea \ket{I}&=&\ket{0}\otimes \ket{vac} \nn \\
\ket{J}&=&\ket{1}\otimes \ket{vac},
\eea
where $\ket{0}$ and $\ket{1}$ are the two internal levels of the ion which form the qubit, and $\ket{vac}$ is the phonon vacuum state (i.e. every mode has zero occupation number); for a ``U'' pulse,
\bea
\ket{I}&=&\ket{0}\otimes \ket{\{1,0,\ldots,0\}} \nn \\
\ket{J}&=&\ket{1}\otimes \ket{vac} ,
\eea
where $\ket{\{1,0,\ldots,0\}}$ is the state in which the first phonon mode has occupation number $1$, and all the others have occupation number zero.  Pulses of these two types for which $\theta$,defined in eq.~(\ref{thetadef}), takes the vales $\pi/2, \pi$ and $2\pi$ can be combined to constitute a quantum logic gate~\cite{CZ}.

\section{Cooling, Quantum Logic and Readout by Raman Transitions}
\setcounter{equation}{0}

We will now discuss  how the various laser interactions described theoretically in the previous section can used to perform the operations required for quantum computing.  These operations can be divided into four parts, namely Doppler cooling of the trapped ions to produce a crystal (this does not, in fact, require Raman transitions); sub-Doppler cooling to reduce the crystal to the quantum ground state of its motion, thereby preparing the quantum register in the required initial state; quantum logic operations; and read-out of final quantum state.

\subsection{Doppler cooling}
Use of standard Doppler cooling~\cite{Wineland:79} is necessary in order to crystallise the trapped ions into the the required string configuration with a temperature small enough to allow sub-Doppler cooling.  Doppler cooling is done as follows: photons from a laser at 397~nm, detuned to a slightly longer wavelength (i.e. to the red) of the $4\,^{2}{\rm S}_{1/2}$ to $4\,^{2}{\rm P}_{1/2}$ transition are absorbed by the ion, whose momentum is reduced by the consequent recoil; this process is followed by spontaneous emission, which causes the ion to recoil in a random direction.  Thus the whole process has the net effect of, on average, reducing the momentum of the ion along the direction of the cooling beam.

Because the upper state of the 397~nm resonance, the $4\,^{2}{\rm P}_{1/2}$ state, can also decay to the long lived $3\,^{2}{\rm D}_{3/2}$ state, another laser at 866~nm must also be employed to pump out any population which might get trapped there.  (see Fig.  3.1).

The lowest temperature attained via Doppler cooling, which is limited by the random recoil of the ions during spontaneous emission, is given by $T_{Dop}=\hbar/2 K_{B} \tau_{PS}$ (where $\tau_{PS}$ is the radiative lifetime of the cooling transition and $K_{B}$ is Boltzmann's constant)~\cite{lascollim}.  For ${\rm Ca}^{+}$, $T_{Dop}=4.96\times 10^{-4}$K (which, for a trapping frequency of $\omega_{x}=(2\pi)\times$500~kHz, corresponds to an average center-of-mass phonon number of $\bar{n}=K_{B}T_{Dop}/\hbar \omega_{x}$ = 20.7).  At this temperature the trapped ions are crystallised (crystallisation can and does occur above the Doppler limit).

\subsection{Sub-Doppler Sideband Cooling via Raman Transitions}
Once a string of ions has been cooled to the Doppler limit, a different cooling technique must be employed to remove the remaining phonon excitations of the string.  One method of doing this is to perform a U-type red sideband transition between the $4\,^{2}{\rm S}_{1/2}$ state and the $3\,^{2}{\rm D}_{3/2}$ state using a 732 nm laser pulse detuned by the trapping frequency $\omega_{x}$; population in the $4\,^{2}{\rm S}_{1/2}$ ground state would be excited to the $3\,^{2}{\rm D}_{3/2}$ excited state, and, at the same time, one phonon excitation would be annihilated.  A V-type carrier transition at 866~nm could then excite the ion to the $4\,^{2}{\rm P}_{1/2}$ state, from whence it would rapidly decay to the $4\,^{2}{\rm S}_{1/2}$ ground state.  Provided the trapping potential is sufficiently strong, there will be only a small probability of the decay process re-exciting the phonon by recoil (this is the Lamb-Dicke effect, discussed below).  Thus the ion would be returned to its initial state, allowing the process to be repeated until all of the phonons are pumped out.  This is a simplified description of single laser sideband cooling, which does not take into account the Zeeman structure of the various levels.  More details can be found in ref.~\cite{thepaper}.

This method, however, requires high frequency stability of the laser in order to resolve the motional sidebands.  Using Raman transitions for the sideband cooling, the laser stability requirement is dramatically reduced \cite{NISTraman}.  Instead of generating a 732~nm laser with a bandwidth of a few kHz, all that is required in the Raman case is that the relative frequency difference between the Pump and Stokes beams be stable.  Sideband cooling using this method consists of two steps.  First, a ``U'' type Raman transition is carefully tuned to resolve motional sidebands, allowing transitions between the two sublevels of the ground state to be performed {\em while at the same time\/} reducing the phonon number by resolving the appropriate sideband.  Fig.~3.2a shows this transition as proceeding from the $m_J=+1/2$ to the $m_J=-1/2$ sublevels of the ground state.  If there was any population in the $m_J=-1/2$ at the start, then this process would have the opposite effect, i.e. it would add a quantum of motion rather than subtracting one.  For this reason we must also carry out the second step, fig.~3.2b, which involves optically pumping the ion back to the original sublevel of the ground state.  This is achieved by using on-resonance circularly polarised light: any population in the $4\,^{2}{\rm S}_{1/2},\,m_J=-1/2$ sublevel will get excited to the $4\,^{2}{\rm P}_{1/2},\,m_J=+1/2$ sublevel, from where it can decay either to the $4\,^{2}{\rm S}_{1/2}\,m_J=-1/2$ or $m_J=+1/2$ sublevels.  In the former case, it will get re-excited back to $4\,^{2}{\rm P}_{1/2},\,m_J=+1/2$, in the later case it will remain, as the circularly polarised light does not have any effect on the $4\,^{2}{\rm S}_{1/2},\,m_J=+1/2$ population.  After 100 nsec or so (i.e. $\sim$ 10 cycles), over 98\% of the population will end up in the $4\,^{2}{\rm S}_{1/2},\,m_J=+1/2$ sublevel, and so step 1 can be repeated.  In order to maintain a high cooling rate the $3\,^{2}{\rm D}_{3/2}$ state must again be pumped out as it has a chance of becoming populated during the optical pumping process; this will require modulation of the 866~nm laser so that all three of the Zeeman levels of the $3\,^{2}{\rm D}_{3/2}$ can be re-pumped.

The Lamb-Dicke effect will stop the ions from being re-heated during the pump-out process (which involves spontaneous emission and therefore the possibility of heating due to recoil).  Specifically, if the kinetic energy imparted to the ion by the spontaneous emission recoil is less than the energy required to excite a phonon, then the recoil is absorbed by the ion trap apparatus as a whole rather than by the individual ion. Hence, for low phonon numbers, we require that
\be
\frac{{\bf p}^{2}}{2 M}\ll \hbar\omega_{x} ,
\label{lamdicone}
\ee
where ${\bf p}=\hbar {\bf k}$ is the momentum due to the recoil (${\bf k}$ being the wavevector of the emitted photon), $M$ is the mass of the ion string ($6.64\times 10^{-26}$kg per ion for Calcium), and $\omega_{x}$ is the angular oscillation frequency of the ion trap.  The inequality (\ref{lamdicone}) can be re-written as follows:
\be
\frac{\hbar k^{2}}{2 M \omega_{x}}\equiv\eta^{2} \ll 1,
\label{lamdictwo}
\ee
where $\eta$ is the Lamb-Dicke parameter, a dimensionless parameter characterizing the coupling between the laser and the motional degrees of freedom of the ion.  The value of $\eta$ depends on the transition involved; its effective value for Raman transitions is discussed in Appendix B. This simplified explanation using classical mechanics is, of course, not valid for the quantum case; a thorough analysis shows that the probability of a recoil phonon being created during the spontaneous emission is approximately equal to the square of Lamb-Dicke parameter, $\overline{n} \eta^{2}$, where $\overline{n}$ is the average phonon number.  Hence for small values of $\eta$ and low phonon numbers, the probability of reheating will be small.  Since $|{\bf k}|=2\pi/\lambda$, where $\lambda$ is the wavelength of the emitted photon (i.e. 397~nm), eq.~(\ref{lamdictwo} implies that the trapping frequency $\omega_{x}$ must satisfy the inequality $\omega_{x}\gg \overline{n} (2\pi) 31.6 \rm{[kHz]}/N$ (where $N$ is the number of ions), in order for sideband cooling to be efficient.

\subsection{Quantum Logic}

Quantum logic operations can be performed using Raman transitions.  As mentioned above, two types of operation are required: V-type carrier transitions (fig.3.3(a)) and U-type red sideband transitions (fig. 3.3(b)).  Selection between U-type and V-type pulses can be made by varying the relative detuning between the Stokes beam and the Pump beam.  The manner in which these pulses can be combined to realise quantum logic gates is described in detail by Cirac and Zoller in their seminal paper~\cite{CZ}, and there is no need to repeat it here.

As mentioned above, the $\ket{1}$ state has a very low probability of spontaneous emission to the $\ket{0}$ state.  Furthermore, because there is only a very small probability of exciting the $4\,^{2}{\rm P}_{1/2}$ level, there is only a very small chance of decays to the $3\,^{2}{\rm D}_{3/2}$ level, and so a pump-out laser at 866~nm is not required during quantum logic operations.

One problem in performing quantum logic operations in this manner is the auxiliary state $\ket{aux}$ required to perform the Cirac-Zoller CNOT gate.  As mentioned in the introduction, $4\,^{2}{\rm S}_{1/2}$ the ground state of $^{40}{\rm Ca}^{+}$ has only two sublevels, which are employed as the logical $\ket{0}$ and $\ket{1}$ states; a third level is required for $\ket{aux}$.  Four possibilities exist.  The first is to use one of the sublevels of the other states~\cite{blatt}:
if we use a P state sublevel, there is a danger of spontaneous emission while the auxiliary rotation is being performed; if we use a D state sublevel, the quantum logic scheme will require either phase-locked lasers 397~nm and 866~nm to perform Raman transitions between the S to D levels, via the P level, or the construction of a narrow bandwidth 732~nm laser.

The second possibility is to use one of the other sidebands (i.e. a sideband {\em not} associated with the CM mode but one associated with the breathing mode, for example)~\cite{Stevens:98}.  However, if the sidebands are close together in frequency space that this method will not be particularly easy.

The third way is to use an isotope of calcium which has hyperfine structure.  This requires an isotope with non-zero nuclear spin, the only stable one being $^{43}{\rm Ca}$ (0.135\% natural abundance) with nuclear spin 7/2.  Alternatively radio-isotopes can be used (see Table 1.), of which $^{41}{\rm Ca}$ or $^{45}{\rm Ca}$ appear to be the most promising candidates.  Although these isotopes are very scarce, and therefore expensive to purify, it should be noted that ion traps lend themselves rather well to use as mass filters, and so can be used for {\em in situ} isotope separation.  DeVoe and Kurtsiefer at IBM Almaden are performing such isotope purification in their ${\rm Ba}^{+}$ experiment, although the natural abundance of the desired isotope of barium is considerably larger than that of calcium~\cite{DeVoePC}.

The fourth possibility is to change the quantum computational paradigm so that an auxiliary level is no longer needed.  Wineland's group at NIST Boulder has described such an alternative to Cirac and Zoller's quantum logic scheme which has the added advantage of requiring fewer pulses~\cite{NISTsimpgate}.  However this scheme does require precise control of laser beam directions and trap frequencies, so that exact and rather large values of the effective Lamb-Dicke parameter are attained.

\subsection{Readout}

An efficient readout scheme is the one aspect of quantum computation in which ion traps are far superior to any of the other proposed technologies.  The basic idea is to use a ``cycling transition'' in which population from one of the two quantum levels of each qubit, $\ket{1}$ say, is repeatedly excited by a laser to some short-lived state.  This short-lived excited state then decays back to the original state $\ket{1}$, resulting in observable fluorescence, without disturbing any of the population in the other quantum state $\ket{0}$. This technique was used successfully in the breakthrough quantum gate experiment performed at Boulder and reported in 1995~\cite{NISTgate}.

In the scheme under discussion here, the readout can be achieved by a two-step process as follows.  First any population in the $4\,^{2}{\rm S}_{1/2},\,m_J=+1/2$ sublevel (i.e. the $\ket{1}$ state of each qubit) is transferred to the $3\,^{2}{\rm D}_{5/2}$ level by a pulse of a 729~nm laser (Fig. 3.4(a)). This laser does {\em not} need to resolve individual motion sidebands of this transition, and so it does not have a particularly stringent bandwidth requirement. We shall refer to this as the ``shelving transition''.

Secondly, the $4\,^{2}{\rm S}_{1/2},\,m_J=-1/2$ to $4\,^{2}{\rm P}_{1/2},\,m_J=-1/2$ transition is excited using the 397~nm laser (Fig. 3.4(b)).  If there is any population in the $4\,^{2}{\rm S}_{1/2},\,m_J=-1/2$ state (i.e. the $\ket{0}$ state of the qubit), there will be fluorescence which we can observe.  Note that there will be decay from $4\,^{2}{\rm P}_{1/2},\,m_J=-1/2$ to the $3\,^{2}{\rm D}_{3/2},\,m_J={-3/2, -1/2\, \&\, +1/2} $ states, which will trap the populations.  Therefore, just as was the case during Doppler cooling, these states must be pumped out using 866~nm light.

If one were using $^{43}{\rm Ca}^{+}$ ions, one might consider an alternative scheme based directly on that successfully used by the Boulder group~\cite{NISTgate}.  The $4\,^{2}{\rm S}_{1/2},\,F=3, m_F=+3$ and the $4\,^{2}{\rm S}_{1/2},\,F=4, m_F=+4$ hyperfine sublevels would form the $\ket{0}$ and the $\ket{1}$ states, respectively.  The cycling transition would then consist of right-circularly polarised light tuned to excite population in $\ket{1}$ to the $4\,^{2}{\rm P}_{3/2},\,F=5, m_F=+5$ state.  If population excited to this level were to decay directly to the ground state, then, because of the electric-dipole selection rules, it could only return to the $\ket{1}$ sublevel, resulting in fluorescence which could be observed.  However, the $4\,^{2}{\rm P}_{3/2}$ level can also decay to both the $3\,^{2}{\rm D}_{3/2}$ and the $3\,^{2}{\rm D}_{5/2}$ metastable levels.  Since re-pumping population trapped in these levels in a manner that avoids any false signals is a very complicated problem, this readout scheme can be effectively ruled out.

Another alternative has been suggested by Steane and his collaborators at Oxford~\cite{Stevens:98}.  Instead of performing the shelving transition, they propose to optically pump the population of the $4\,^{2}{\rm S}_{1/2},\,m_J=+1/2$ state via the $4\,^{2}{\rm P}_{3/2},\,m_J=+3/2$ state to the $3\,^{2}{\rm D}_{5/2}$ level, followed by resonance fluorescence on the $4\,^{2}{\rm S}_{1/2}$ to $4\,^{2}{\rm P}_{1/2}$ 397~nm transition (with the usual 866~nm laser to avoid trapping population in the $3\,^{2}{\rm D}_{3/2}$ level). However the $4\,^{2}{\rm P}_{3/2},\,m_J=+3/2$ state can decay to both the the $3\,^{2}{\rm D}_{5/2}$ and the $3\,^{2}{\rm D}_{3/2}$ levels, so that the optical pumping scheme has a small (10$\%$) chance of causing a false signal.  Although this is a drawback, this scheme does have the advantage that a laser in the 729-732~nm range is not required.

Once the readout has been accomplished, the population shelved in the $3\,^{2}{\rm D}_{5/2}$ level has to be removed.  This can be achieved either by repeating the shelving pulse with the 729~nm laser, or by using two lasers at 866~nm and 854~nm to pump population from the metastable state D-states to the ground state via the $4\,^{2}{\rm P}_{3/2}$/ $4\,^{2}{\rm P}_{1/2}$ levels.  The second technique is quicker and more reliable, but it does require a third laser wavelength at 854~nm.

\section {Experimental Requirements}

In this section we will discuss the specifications of the apparatus needed to carry out the experiments described above.  A full description of the ion trap, the vacuum system, the associated lasers and optical systems can be found in ref.~\cite{thepaper}.  Here we will confine ourselves to a description of the additional apparatus needed specifically for the stimulated Raman transitions described above.

\subsection{Static magnetic field}

A strong static magnetic field is necessary in order to split the Zeeman sub-levels in frequency space, so that they may be reliably resolved with the available lasers.  The Zeeman structure of the 397~nm $4\,^{2}{\rm S}_{1/2}$ to $4\,^{2}{\rm P}_{1/2}$ transition and of the 866~nm $4\,^{2}{\rm D}_{3/2}$ to $4\,^{2}{\rm P}_{1/2}$ transitions are shown in figs. 4.1(a) and 4.1(b), respectively.  The minimum splittings between component lines are $(2\pi) 0.93 {\rm MHz/Gauss}$ and $(2\pi) 0.19 {\rm MHz/Gauss}$ respectively, while the radiative line widths are $129 {\rm MHz}$ and $10.6 {\rm MHz}$, respectively.  Thus a magnetic field strength of 200 Gauss will be sufficient for the Zeeman splitting of both lines to be well resolved.

The stability requirements for the static magnetic field are important. Variations in the magnetic field during quantum operations will mean that there is an error in the effective Rabi frequency and so there will be an over- or under-rotation during the logic operations. The error in angle of rotation is
\be
\delta\theta \approx -\theta \frac{\delta B}{B}.
\ee
Thus if we require a angular accuracy of say $\delta\theta/\theta \sim 10^{-6}$ (corresponding to an error rate below the accuracy threshold for fault-tolerant quantum computing~\cite{Laf}) with each operation taking $\sim$10 $\mu$sec, and we have a magnetic field of 200 Gauss, then the rate of change of the field must be less than 20 Gauss per second. The magnetic field may be stabilised by placing a Hall probe in the field and using it as a sensor for a feedback loop to the Helmholtz coils.

\subsection{Lasers}

\subsubsection{Power}
The power requirements of the different laser tasks needed for quantum computing are approximately given by the following formulas (for derivation, see refs.~\cite{thepaper, DFVJ} and also see Appendix B for Raman transitions):
\be
P\approx\left\{
\begin{array}{ll}
\frac
{\D \Omega^{2} h c d^{2} \tau}
{\D \lambda^{3}}&
\mbox{(single laser, V pulse)}\\
&\\
\frac
{\D \Omega \Delta h c d^{2} \tau}
{\D \lambda^{3}}&\mbox{(Raman, V pulse)}\\
&\\
\frac
{\D \Omega \Delta h c d^{2} \tau}
{\D \lambda^{3}}
\left(\D \frac{\sqrt{N}}{\D \eta}\right)&\mbox{(Raman, U pulse)}\\
&\\
\frac
{\D 4 h c d^{2} }
{\D \tau \lambda^{3}}&
\mbox{(single laser, saturation power).}\\
\end{array}
\right.
 \ee
In these expressions, $\Omega$ is the effective Rabi  frequency for the process, $h$ is Planck's constant, $c$ is the speed  of light, $d$ is the $1/e^{2}$ diameter of the intensity pattern of the laser in the vicinity of the ion (we have assumed a  $\mbox{TEM}_{00}$ Gaussian beam), $\lambda$ and $\tau$ are the wavelength and radiative lifetime of the transition, respectively, $\Delta$ is the Raman detuning, $\eta$ is the Lamb-Dicke parameter and $N$ is the number of ions.

For purposes of estimating power, we assume that we wish to obtain Rabi frequencies equal to $(2\pi) 1.0~ {\rm MHz}$ for all the operations.  We require tight focusing only for the 397~nm laser during quantum logic operations; the other two lasers (866~nm and 729~nm), and the 397nm laser during Doppler cooling and readout, do not need to address individual ions; in fact it is more convenient for them to have a broad focal pattern with intensities equal for all ions in the trap.  Therefore we assume a $d = 10 \mu$m focal spot diameter for the 397~nm laser during logic operations and $d = 100 \mu$m focal spot diameters for the other lasers.  The Lamb-Dicke parameter was assumed to be $\eta=0.1$, the number of ions to be $N=10$ and the Raman detuning to be $\Delta=10$~GHz.

The 397~nm laser will be the workhorse of a Raman-based Calcium ion quantum computer.  It will be required to perform Raman transitions (of both the U and V types) on single ions for quantum logic, U-type transitions on all of the ions (with a broad focal spot size) for sideband cooling, and to be used to saturate the $4\,^{2}{\rm S}_{1/2}$ to $4\,^{2}{\rm P}_{1/2}$ transition during the Doppler cooling and the readout phases.  We calculated the power required for all of these operations and found that the U-type Raman transition during sideband cooling required the most power (which is the figure given in Table 2.)  The 866~nm laser is needed for pump out, acting on all of the ions (and therefore having a broad focal spot); its power is the saturation power for that transition.  The 729~nm laser is used for a single V-type $\pi$ -pulse shelving transition during readout. Because the $4\,^{2}{\rm S}_{1/2}$ to $3\,^{2}{\rm D}_{5/2}$ transition is dipole forbidden, this laser has the highest power requirements.

\subsubsection{Optical system}

In order to fulfill momentum conservation and quantum selection rules, the Raman and Stokes beams are required to have different polarisations and propagation directions, i.e. each ion must be physically addressed by two laser beams, each with slightly different frequency, polarisation, and direction.  This complicates the ion addressing optics, but not unduly so: Figure 4.2 shows a simple experimental layout for producing appropriate Pump and Stokes beams. The arrangement shown has the disadvantage that the effective wavevector $\bf{k}_{s}-\bf{k}_{p}$ has a component transverse to the weak axis of the trap, and so care must be taken to avoid the creation of transverse phonons when using it.  The alternative is to use a optical arrangement in which the pump and Stokes beams are no longer co-propagating, which requires duplicate optical addressing system for the two beams.  However, only one of the pump or Stokes beams needs to be tightly focused to address the ions; the other beam can have a broad focal spot, acting on all of the ions simultaneously (and resulting in an A.C. Stark shift of the $4\,^{2}{\rm S}_{1/2}$ to $4\,^{2}{\rm P}_{1/2}$ 397~nm transition); transitions would only occur when the second, tightly focused laser beam is acting on the ion~\cite{DeVoePC}.

\subsubsection{Laser Stability}

The bandwidths of the 397~nm and 866~nm lasers are determined by the strength of the magnetic field: The Zeeman splitting of the transitions, discussed above, must be resolvable in frequency by the lasers.  There is no need for lasers to be narrower than the natural widths of the lines (which we have used as the bandwidths cited in Table 2.

For the 729~nm laser needed for the shelving pulse during readout, the stability is determined by the accuracy of the $\pi$-pulse.  For a on-resonance laser, all of the population in the lower level is transferred to the upper level by the $\pi$-pulse.  However, if the laser is detuned by an amount $\Delta$ because of its finite bandwidth, there will be a probability $(\Delta/\Omega_{0})^{2}$ of remaining in the $\ket{1}$ state (and consequently causing a false readout signal).  Here $\Omega_{0}$ is the single laser Rabi frequency, which may be calculated using formulas given in ref.~\cite{DFVJ}.  For $\Omega_{0}=(2\pi) 1.0{\rm MHz}$, a detuning of $\Delta = (2\pi) 100 {\rm kHz}$ gives a $0.3\%$ chance of this happening; thus a bandwidth of $\sim (2\pi) 100 {\rm kHz}$ should be sufficient for the 729~nm laser.  Such bandwidth are now standard.  A more accurate readout can be achieved by either increasing the 729~nm laser power (thereby increasing $\Omega_{0}$) or by decreasing the bandwidth.

\subsubsection{Pump out laser}

As mentioned in section 3, the Zeeman splitting of the 866~nm transition (shown in fig.  4.1) will require some form of frequency modulation of the pump-out laser at that wavelength.  For a 200 Gauss magnetic field, the amount of modulation is roughly 280 MHz.  One way to achieve this is with two electro-optic modulators (EOMs) in series, one generating frequency sidebands at 18.6 MHz (i.e. $\Delta \omega /15$ ), the other generating sidebands at 224 MHz (i.e. $ 12\Delta \omega /15$).

\subsubsection{Radio-frequency oscillator}

Frequency differences between the pump and Stokes beams of several GHz can be achieved with commercial radio-frequency (RF) sources and electro-optic modulators.  The RF signal needs be stable, the stability required being directly dependent on the required accuracy of the logic operations.  If we set this as 1 part in $10^{6}$ we need an RF bandwidth of about 1 kHz.  Thus the phase stability of commercial RF sources is more than adequate for elementary quantum logic experiments.

Table 2 shows the laser required characteristics calculated on the basis of the considerations discussed above.  Note that the calculated laser intensity is for the focal region and so does not take into account losses due to various effects in the optical system.

\section {Conclusions}

This paper has presented in detail a scheme for performing quantum logic operations with trapped calcium ions using stimulated Raman transitions.  Speaking from our own experience in establishing an ion trap quantum computing experiment~\cite{thepaper}, the most challenging single experimental problem is the generation of a large number of powerful cw laser beams at a variety of wavelengths and with very stringent stability requirements.  This is particularly the case when one attempts a ``single laser'' transition scheme, exploiting the Rabi oscillations between two levels to perform the necessary operations.  It was this difficulty which was the principal motivation for the approach we have considered here: the adoption of Raman type laser transitions means that absolute phase stability of lasers is no longer a critical issue.  However some of the resultant experimental arrangements are more complex.  In particular, there are stringent requirements for the stability of the static magnetic field.  As experimental quantum computing is without a doubt a very challenging problem, one must have as many technological options as possible if one hopes to meet with success.

\vspace{1cm} \noindent {\em Acknowledgments}.

The authors wish to thank Rainer Blatt, Ralph DeVoe, Richard Hughes, Christian Kurtsiefer, Christof Naegerl, Andrew Steane, Dale Tupa and David Wineland for useful conversations.  This work was funded by the National Security Agency.

\newpage
\appendix
\section{Adiabatic elimination of the upper levels}
\setcounter{equation}{0}

The elimination of the upper state populations can be carried out as follows. The formal solution of eq.~(\ref{bdot}) is:
\be
b_{k}(t)=b_{k}(0)+\frac{\ci}{2}\sum_{j'}\sum_{a}
\Omega^{(a)\ast}_{j'k}
\int^{t}_{0}a_{j'}(t')\exp\left(\ci\Delta^{(a)}_{j'k}t'\right)dt' .
\ee
If one now makes the Markov approximation (i.e. assume that $a_{j'}(t)$ is slowly varying so that it can be factored out of the integral) one obtains the following approximate expression for the amplitude in the upper manifold:
\be
b_{k}(t)\approx b_{k}(0)+\sum_{j'}\sum_{a}
\frac{\Omega^{(a)\ast}_{j'k}}{2\Delta^{(a)}_{j'k}}
a_{j'}(t)\left[\exp\left(\ci\Delta^{(a)}_{j'k}t\right)-1\right]
\label{druitt}.
\ee
We can assume that the initial amplitude in any of the upper manifold of states is zero, i.e. $b_{k}(0)=0$.  Since all of the states of the upper manifold have a dipole allowed transition to one or more states of the lower manifold, it is reasonable to assume that any residual population in the upper manifold will rapidly decay due to spontaneous emission.  On substituting from eq.~(\ref{druitt}) into the equation of motion for $a_{j}(t)$, eq.~(\ref{adot}), and neglecting a rapidly
oscillating term proportional to $\exp\left(-\ci\Delta^{(a)}_{jk}t\right)$, one obtains the formula eq.~(\ref{adot2}) for the amplitudes in the lower manifold.

\section{Calculation of the effective Rabi Frequency in terms of atomic constants}
\setcounter{equation}{0}

From eqs.~(\ref{parker}) and (\ref{hulme}), we obtain the following formula for the effective Rabi frequency:
\be
\Omega_{eff}=\sum_{k}
\frac{
\bra{i}\hat{\bf{d}}\cdot\bep^{(p)}
E^{(p)}(\hat{\bf{r}})\ket{k}
\bra{k}\hat{\bf{d}}\cdot\bep^{(s)\ast}
E^{(s)\ast}(\hat{\bf{r}}) \ket{j}
}{4\hbar^{2}
\{\omega_{k}-(\omega_{i}+\omega_{j}+\omega_{p}+\omega_{s})/2\}} .
\ee
In order to calculate this quantity in terms of atomic constants etc., it is necessary to return to the representation of the states in terms of tensor products of internal states with external (vibrational) states, vis:
\bea
\ket{i}&\rightarrow&\ket{0}\otimes\ket{\{m\}} \nn \\
\ket{j}&\rightarrow&\ket{1}\otimes\ket{\{n\}} \nn \\
\ket{k}&\rightarrow&\ket{\lambda}\otimes\ket{\{\ell\}} , \eea
where $\ket{0}$ and $\ket{1}$ are two internal states of the lower manifold (i.e. the two states that store the information of the qubit of the quantum computer), $\ket{\lambda}$ denotes an internal state in the upper manifold, and $\ket{\{m\}}$, $\ket{\{n\}}$ and $\ket{\{\ell\}}$ denote the state of the external degrees of freedom of the ion (i.e. the vibrational state) specified by a set of occupation numbers for the different modes, as discussed above.  Note that we have used a simplified notation for the internal states, the single symbol (0, 1, or $\lambda$) denoting the set of quantum numbers $\gamma, J, m_J$ needed to specify the internal state completely.

Using this notation we obtain the following formula for the effective Rabi frequency:
\be
\Omega_{eff}=\sum_{\lambda}\sum_{\ell}
\frac{
\bra{0}\hat{\bf{d}}\cdot\bep^{(p)}
\ket{\lambda}\bra{\lambda}\hat{\bf{d}}\cdot\bep^{(s)\ast}
\ket{1}
\bra{\{m\}}E^{(p)}(\hat{\bf{r}})\ket{\{\ell\}}
\bra{\{\ell\}}E^{(s)\ast}(\hat{\bf{r}}) \ket{\{n\}}
}{4\hbar^{2}
(\Deltabar_{\lambda}-\omega^{(ph)}_{\ell})} ,
\ee
where
$\Deltabar_{\lambda}=\om{\lambda}-(\om{0}+\om{1}+\om{p}+\om{s})/2$
and
$\omega^{(ph)}_{\ell}=\om{x}[\sqrt{\mu_{1}}\{\ell_{1}-(m_{1}+n_{1})/2\}+
\sqrt{\mu_{2}}\{\ell_{2}-(m_{2}+n_{2})/2\}+\ldots
\sqrt{\mu_{N}}\{\ell_{N}-(m_{N}+n_{N})/2\}]$.
We will make the approximation that $\Deltabar_{\lambda}\gg\omega^{(ph)}_{\ell}$, which is justified in most experimental circumstances, where the detuning $\Deltabar_{\lambda}$ is of the order of GHz and the trap frequency $\om{x}$ is usually is no higher than a few MHz. Further, we will assume that both the laser beams are, to a good approximation, plane waves, i.e.
\bea
E^{(p)}(\hat{\bf{r}})&=&E^{(p)}_{0}\exp\left(-\ci
\bf{k}_{p}\cdot\hat{\bf{r}}\right) \nn \\
E^{(s)}(\hat{\bf{r}})&=&E^{(s)}_{0}\exp\left(-\ci
\bf{k}_{s}\cdot\hat{\bf{r}}\right) .
\eea
Thus, if we use the completeness of the vibrational states $\sum_{\ell} \ket{\{\ell\}}\bra{\{\ell\}}=\hat{I}$, then we get the following expression for $\Omega_{eff}$:
\be
\Omega_{eff}=\sum_{\lambda}
\frac{
\bra{0}\hat{\bf{d}}\cdot\bep^{(p)}\ket{\lambda}
\bra{\lambda}\hat{\bf{d}}\cdot\bep^{(s)\ast}
\ket{1}}
{4\hbar^{2}\Deltabar_{\lambda}}
E^{(p)}_{0}E^{(s)\ast}_{0}
f\left(\{m\},\{n\},\eta\right),
\label{loeb}
\ee
where
\be
f\left(\{m\},\{n\},\eta\right)=\prod_{p}\bra{m_{p}}
\exp[\ci\xi_{p}(\had_{p}+\ha_{p})]\ket{n_{p}}.
\label{deeming}
\ee
In eq.~(\ref{deeming}), $\ket{n_{p}}$ represents the state of the p-th oscillator mode with occupation number $n_{p}$, $\ha_{p}(\had_{p})$ is the annihilation (creation) operator for that mode and, for the s-th ion in the chain, $\xi_{p}=\eta b^{(p)}_{s}/\mu_{p}^{1/4}$, $b^{(p)}_{s}$ being the s-th element of the p-th mode normalised eigenvector (ref.~\cite{DFVJ}, table 2) and, as mentioned above, $\mu_{p}$ is the eigenvalue.  The Lamb-Dicke parameter $\eta$, which is a dimensionless number that characterises the coupling between the laser and the vibrational degrees of freedom of the ion, is given by the formula:
\be
\eta=\left(\bf{k}_{s}-\bf{k}_{p}\right)\cdot\bep_{x}
\sqrt{\frac{\hbar}{2M\om{x}}},
\ee
$M$ being the mass of one of the ions. Note that for stimulated Raman transitions, the Lamb-Dicke parameter depends on the projection of the difference of the two laser's wavevectors onto the x-axis.  Thus by carefully selecting the directions of laser propagation, one has in principle a large degree of control over this coupling.

The matrix elements appearing in eq.~(\ref{deeming}) can be calculated in closed form (see: ref.~\cite{Wineland:79}, section V.A).  The result is
\be
\bra{m}\exp[\ci\xi(\had+\ha)]\ket{n}=
\sqrt{\frac{{\rm min}\{m,n\}!}{{\rm max}\{m,n\}!}}
|\xi|^{|n-m|}L^{|n-m|}_{{\rm min}\{m,n\}}(\xi^{2})
\exp(\xi^{2}/2)\exp(\ci\Phi_{m,n}) ,
\label{lag}
\ee
where
\be
\Phi_{m,n}=\pi[n+{\rm min}\{m,n\}+{\rm sgn}(\xi)(m-n)/2].
\label{lagph}
\ee
In eqs.(\ref{lag}) and (\ref{lagph}), $L^{a}_{n}(x)$ is the Laguerre Polynomial~\cite{GR}, ${\rm sgn}(x)$ is the sign of $x$, and ${\rm min}\{m,n\}({\rm max}\{m,n\})$ is the smaller (larger) of the two integers $m,n$.

For quantum computing we are particularly interested in performing two types of operation, namely the ``V'' type or carrier pulse, for which $\{n\}=\{m\}$ and the ``U'' type or red sideband pulse, for which $\{n\}=1,0,\ldots,0$ and $\{m\}=0,0,\ldots,0$. For these two ideal cases, one can show that
\be
f\left(\{m\},\{n\},\eta\right)=\left\{\begin{array}{ll}
1+O(\eta^{2})&\mbox{(V pulse)}\\
\eta/\sqrt{N}+O(\eta^{2})&\mbox{(U pulse)}
\end{array}
\right.
\ee

The atomic matrix elements appearing in eq.~(\ref{loeb}) can also be calculated.  The result is: \be 
\bra{\nu}\hat{\bf{d}}\cdot\bep^{(a)}\ket{\lambda}= \sqrt{\frac{3 A (2J_{\lambda}+1)}{4 c \alpha k^{3}}} \sum_{q=-1}^{1} \threej{J_{\nu}}{1}{J_{\lambda}}{-m_{\nu}}{q}{m_{\lambda}} \bep_{q}\cdot\bepsa, \ee where $c$ is the speed of light, $\alpha$ is the fine structure constant, the term containing six symbols in round brackets is the Wigner 3j symbol~\cite{RDC}, $\nu$ stands for $0$ or $1$, $a$ stands for $p$ or $s$, $k$ is the wavenumber of the transition (i.e. $k=2\pi/\lambda$, where $\lambda$ is the wavelength) and A the spontaneous decay rate between the upper and lower manifolds (i.e the reciprocal of the radiative decay lifetime) and the vectors $\bep_{q}$ are the usual spherical basis vectors: $\bep_{1}=(-1,\ci,0)/\sqrt{2}, \bep_{0}=(0,0,1), \bep_{-1} = (1,\ci,0)/\sqrt{2}$.

Thus, finally we obtain the following formula for the effective Rabi frequency of Raman transitions:
\be
\Omega_{eff}=\frac{e^{2}A E_{0}^{(p)}E_{0}^{(s)\ast}}
{4 \hbar^{2} c \alpha k^{3}}
\sum_{\lambda}\frac{\beta_{\lambda}}{\Deltabar_{\lambda}}
f\left(\{m\},\{n\},\eta\right) ,
\ee
where $\beta_{\lambda}$ is a dimensionless factor of order unity dependent on the quantum numbers of the upper and lower states and on the polarisations of the two lasers:
\be
\beta_{\lambda}=\frac{3}{4} (2J_{\lambda}+1)
\sum_{q,q'=-1}^{1}
\threej{J_0}{1}{J_{\lambda}}{-M_0}{q}{M_{\lambda}}
\threej{J_1}{1}{J_{\lambda}}{-M_1}{q'}{M_{\lambda}}
\left(\bep_{q}\cdot\bep^{(p)}\right)
\left(\bep_{q'}^{\ast}\cdot\bep^{(s)\ast}\right).
\ee

In order to estimate power requirements, we assume that the pump and the Stokes lasers are of roughly the same intensity, so that $E_{0}^{(p)}E_{0}^{(s)\ast}\approx \left|E_{0}^{(p)}\right|^{2} = 64 P \alpha \hbar /e^{2} d^{2}$, where $P$ is the laser power and $d$ is the $1/e^{2}$ diameter of the laser (we have used ref.~\cite{ME}, eq.(14.5.27) to derive this formula).  Thus if we assume that the
Raman detuning is much larger than the Zeeman splitting of levels in the upper manifold, so that
$\Deltabar_{\lambda}\approx \Delta$, we obtain
\be
\Omega_{eff}=\frac{P \lambda^{3}}{h c \tau d^{2}\Delta}
\Upsilon f\left(\{m\},\{n\},\eta\right),
\ee
where $\Upsilon$ is a factor of order 1.  Using this formula, we obtain the formula for laser power used in section 4.2.

\newpage
\section*{Figures}

\centerline {\includegraphics[width=0.5 \columnwidth]{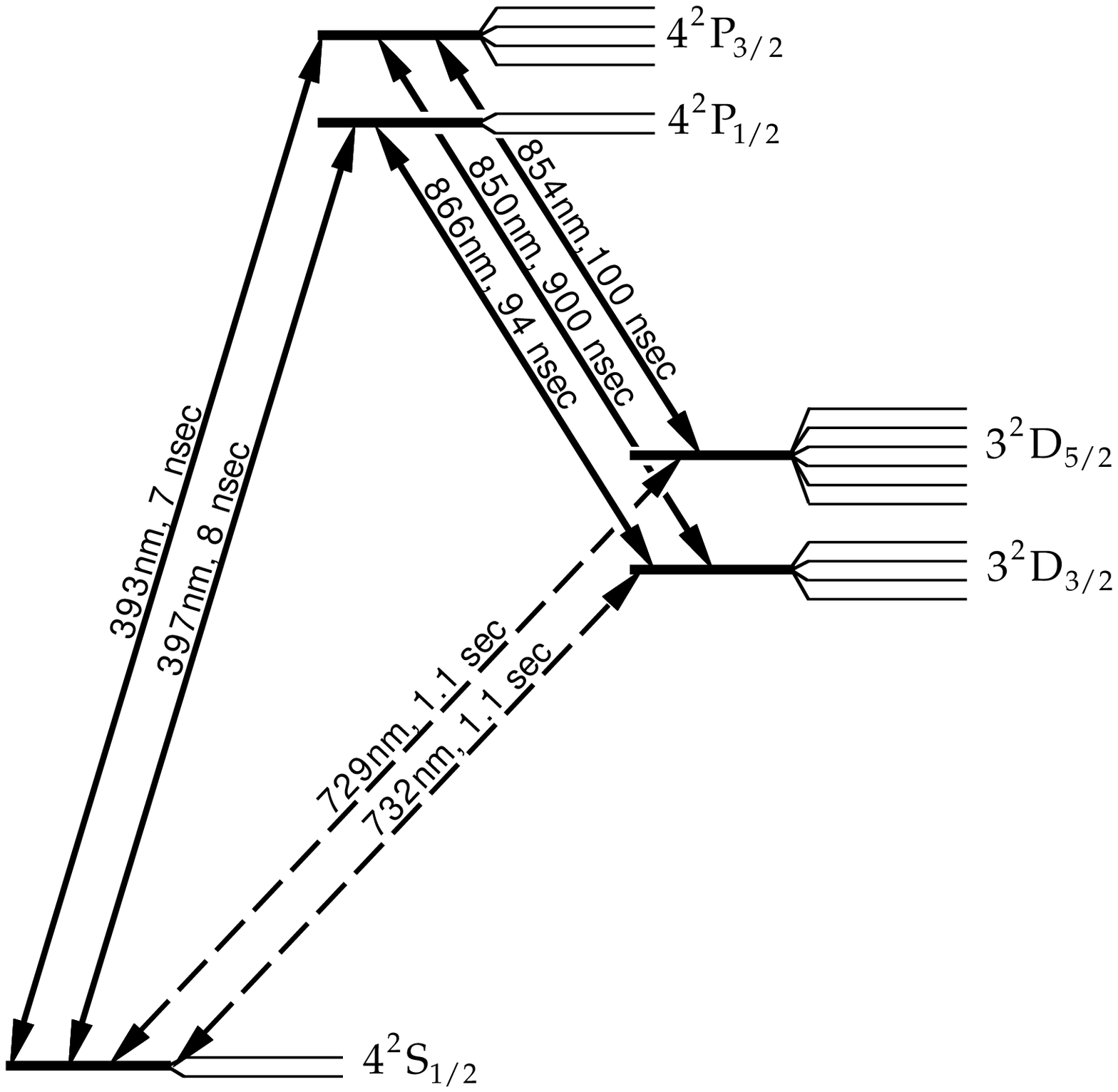}} \noindent
{\bf Figure 1.1.} Low-lying levels of the singly ionised calcium ion, showing all of the Zeeman-split magnetic sublevels, in the absence of hyperfine structure.  The spacing is not to scale.

\centerline {\includegraphics[width=0.5 \columnwidth]{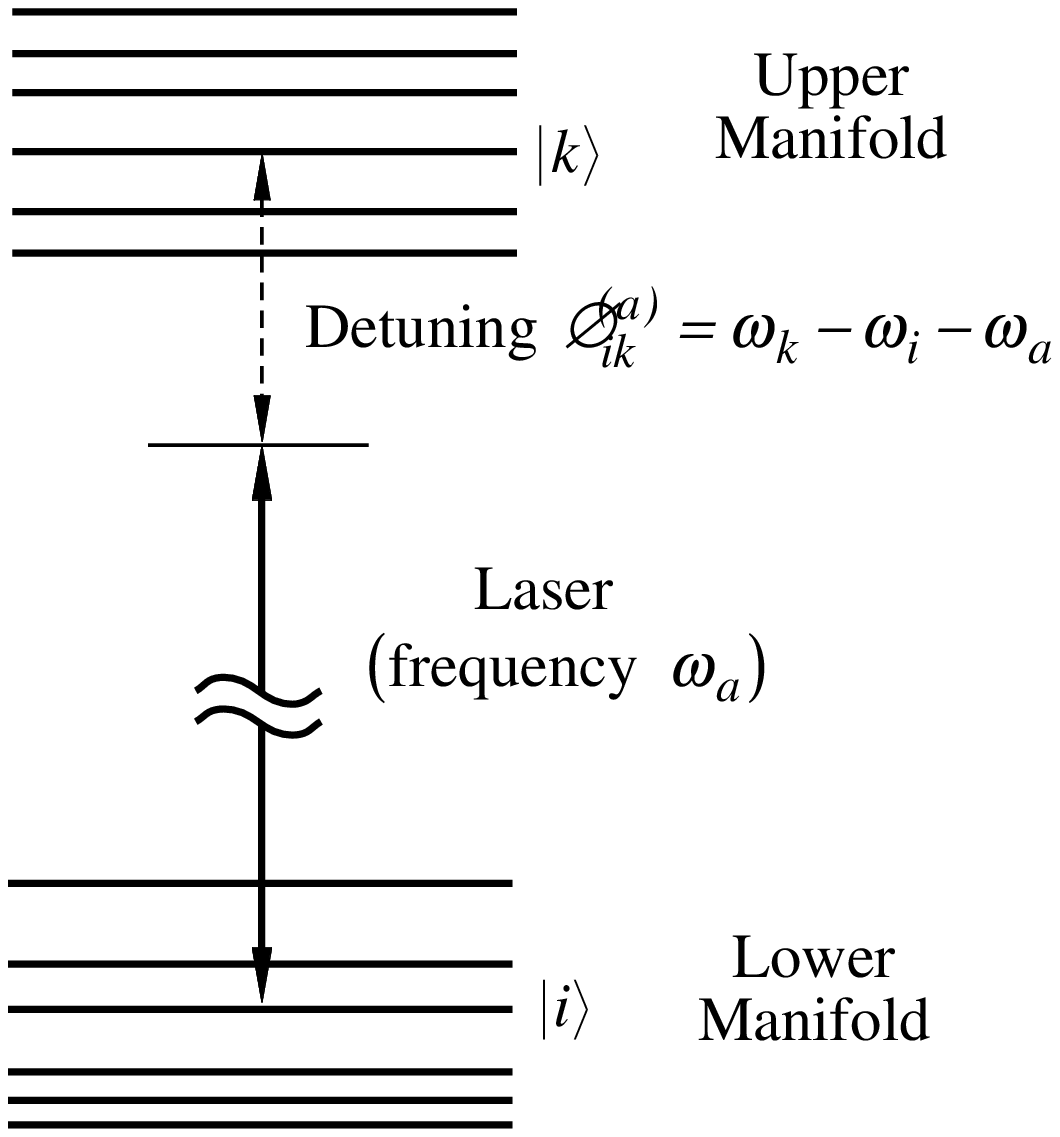}} \noindent
{\bf Figure 2.1.}  Notation used in analysis of Raman transitions in the general case.

\centerline {\includegraphics[width=0.5 \columnwidth]{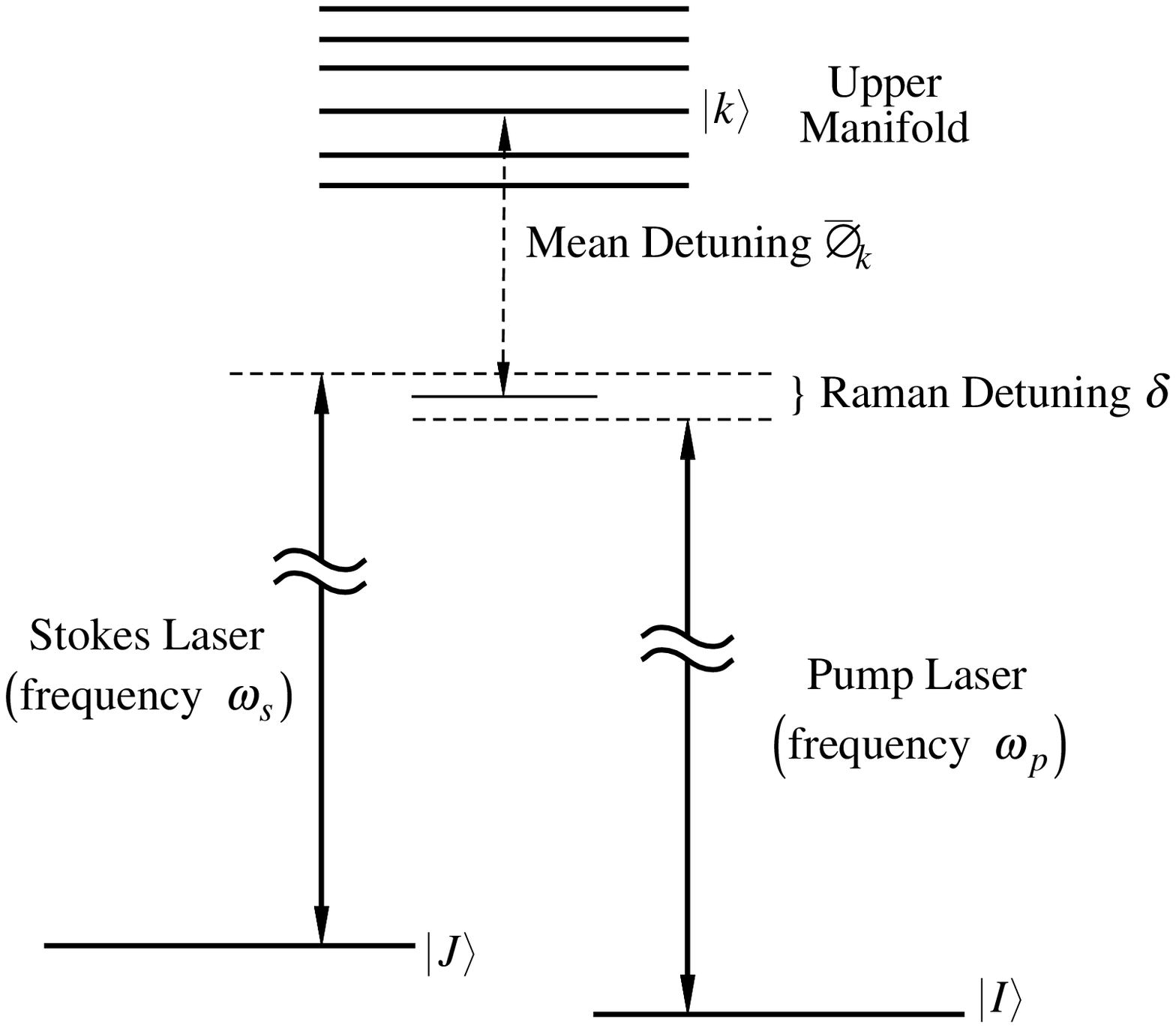}} \noindent
{\bf Figure 2.2.} Notation used in analysis of Raman transitions in two-level systems.

\centerline {\includegraphics[width=0.5 \columnwidth]{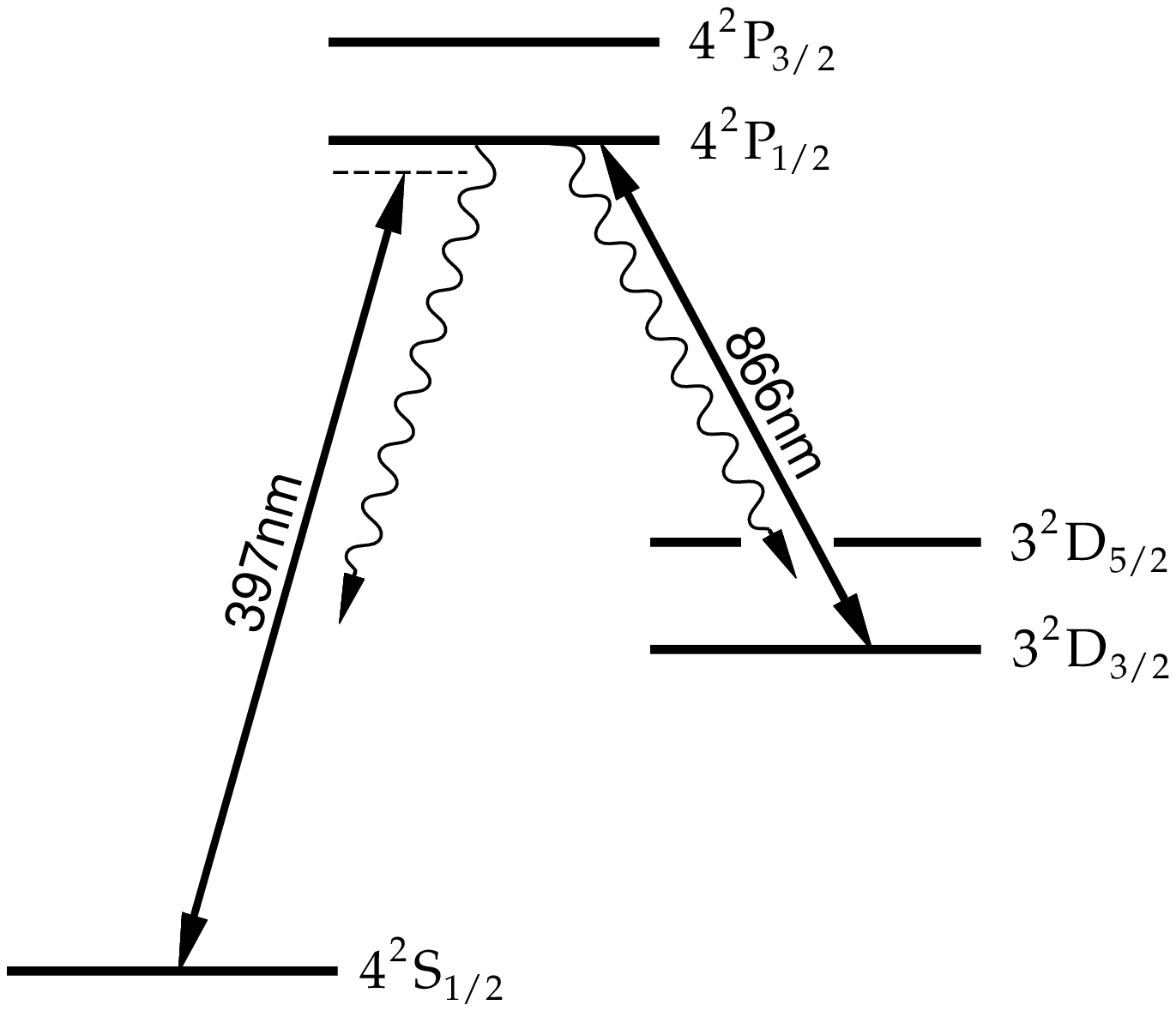}} \noindent
{\bf Figure 3.1.} Doppler cooling.

\centerline {\includegraphics[width=0.4 \columnwidth]{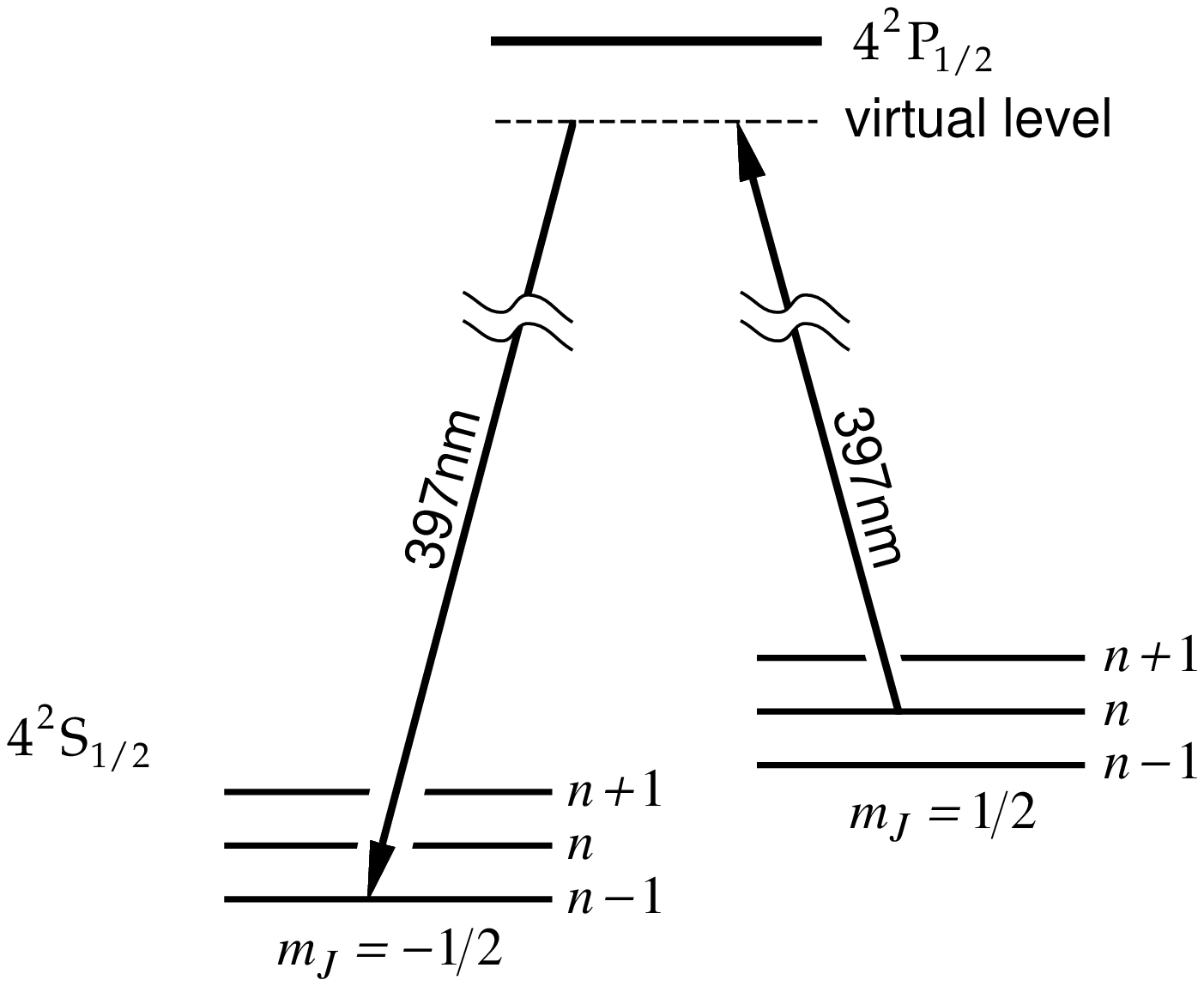}} \noindent
{\bf Figure 3.2(a).} Raman sideband cooling, step 1: removal of one phonon.

\centerline {\includegraphics[width=0.5 \columnwidth]{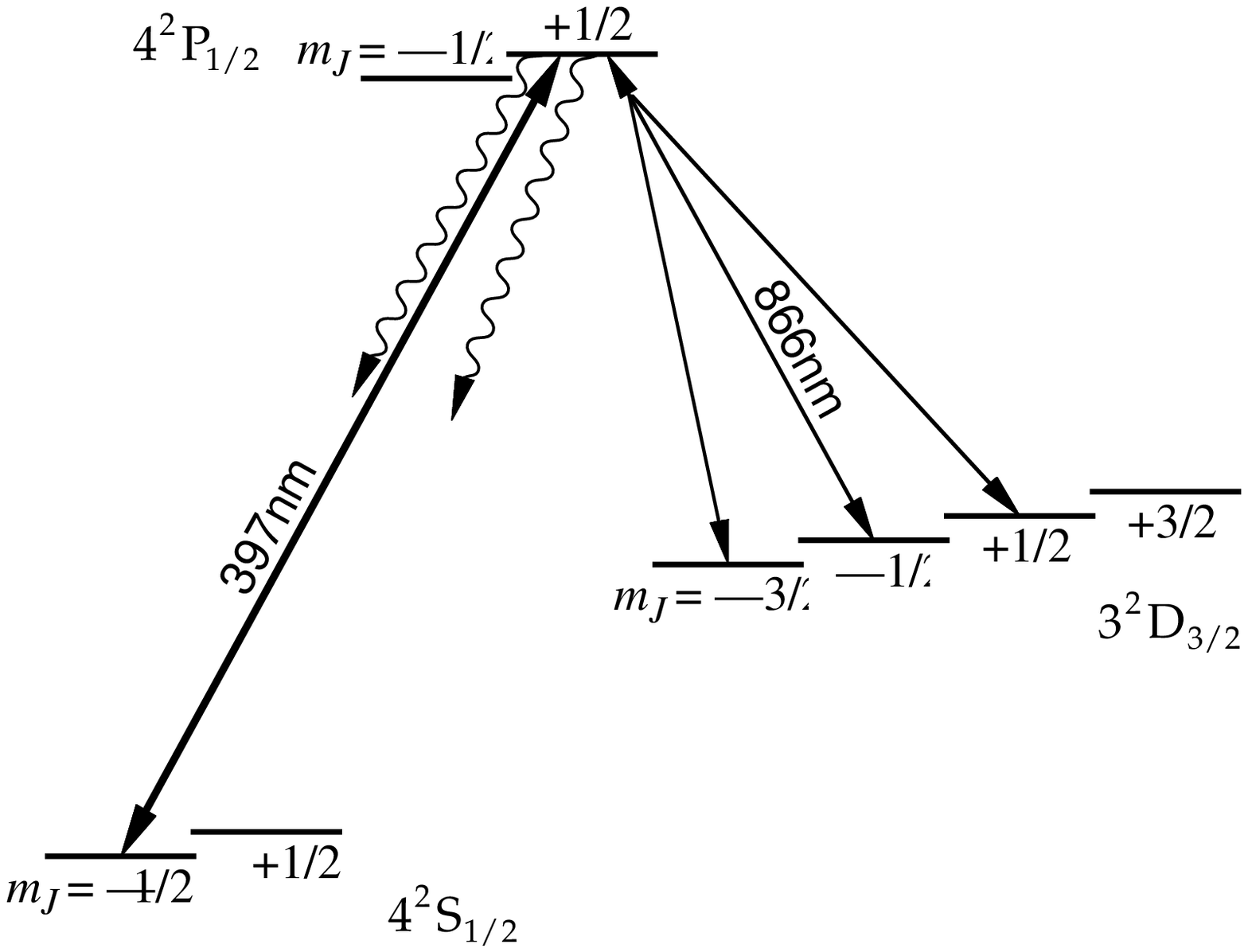}} \noindent
{\bf Figure 3.2(b).} Raman sideband cooling, step 2: recycling by optical pumping.

\centerline {\includegraphics[width=0.5 \columnwidth]{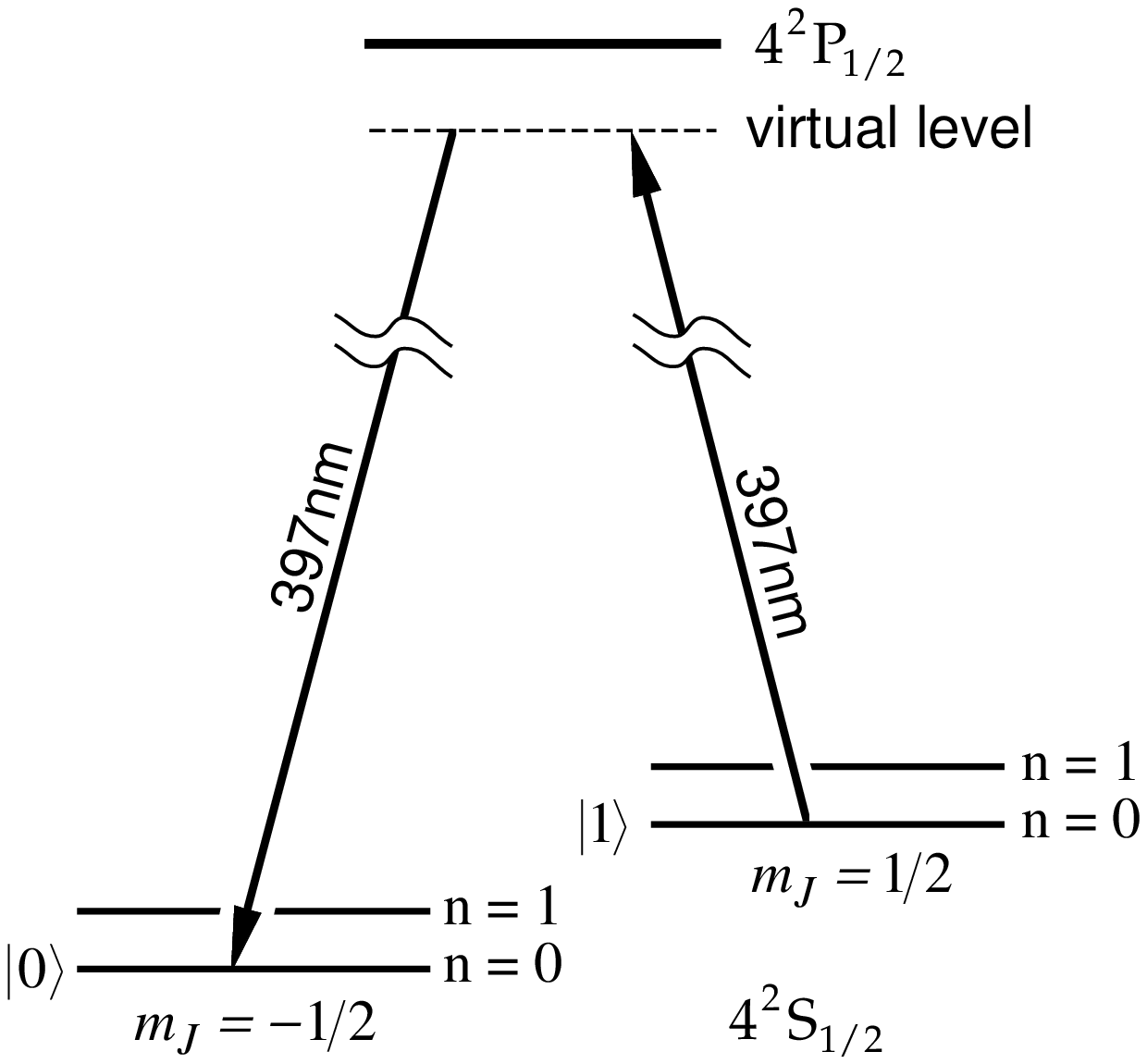}} \noindent
{\bf Figure 3.3(a).} Raman state manipulation: V-Type carrier transitions.

\centerline {\includegraphics[width=0.5 \columnwidth]{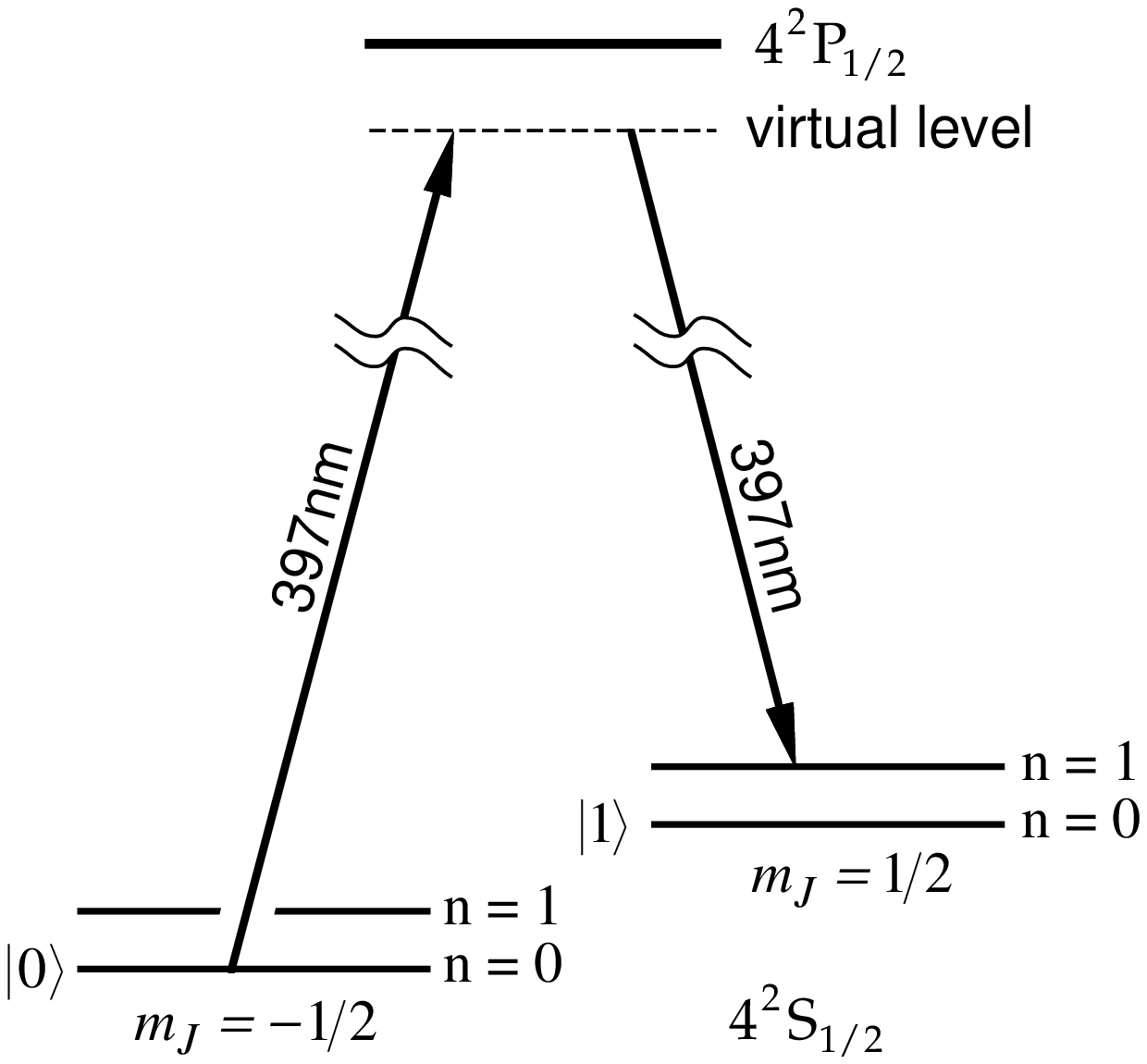}} \noindent
{\bf Figure 3.3(b).} Raman state manipulation: U-Type red-sideband transitions.

\centerline {\includegraphics[width=0.8 \columnwidth]{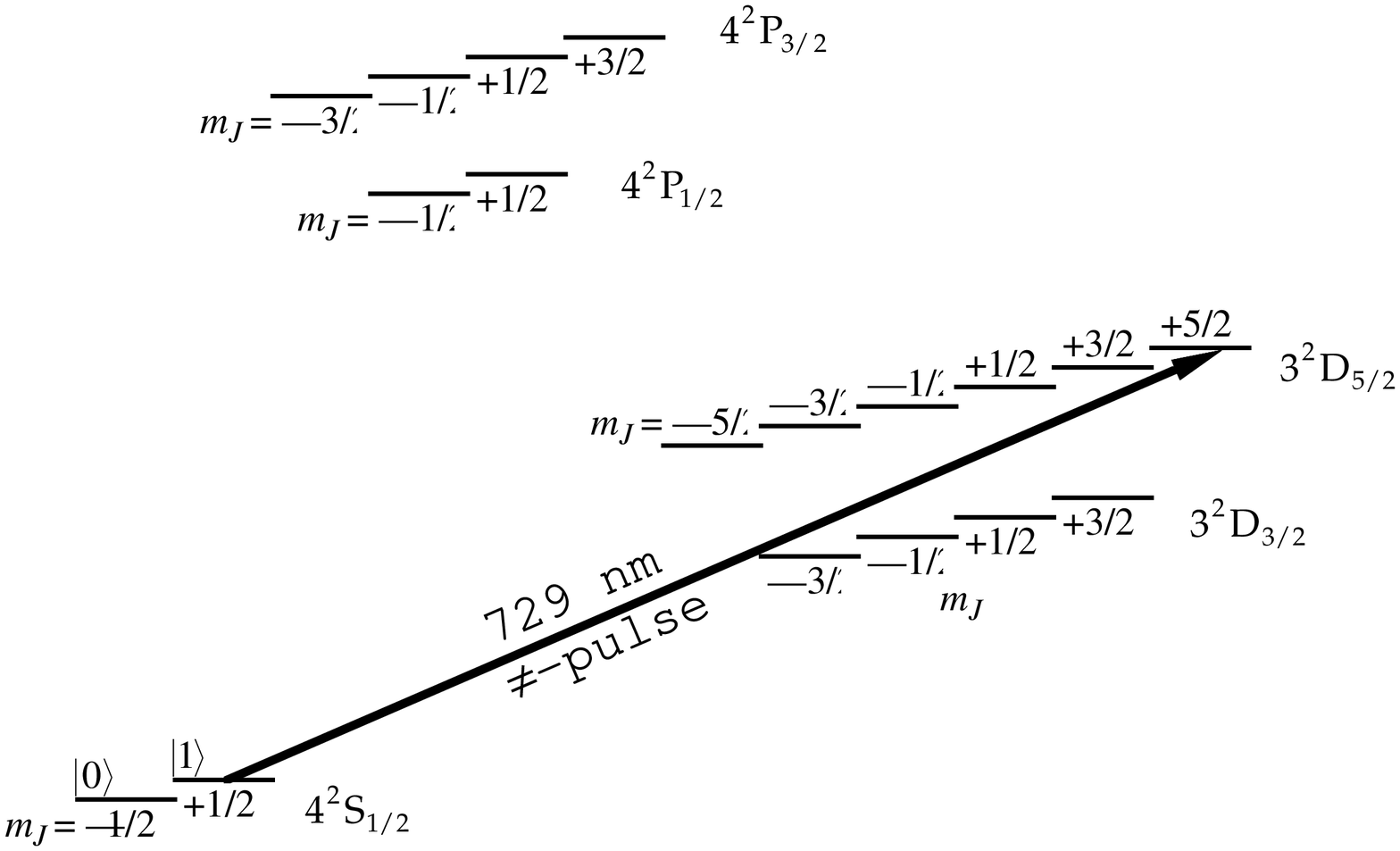}} \noindent
{\bf Figure 3.4(a).} Raman readout, step 1: the ``shelving transition''.

\centerline {\includegraphics[width=0.5 \columnwidth]{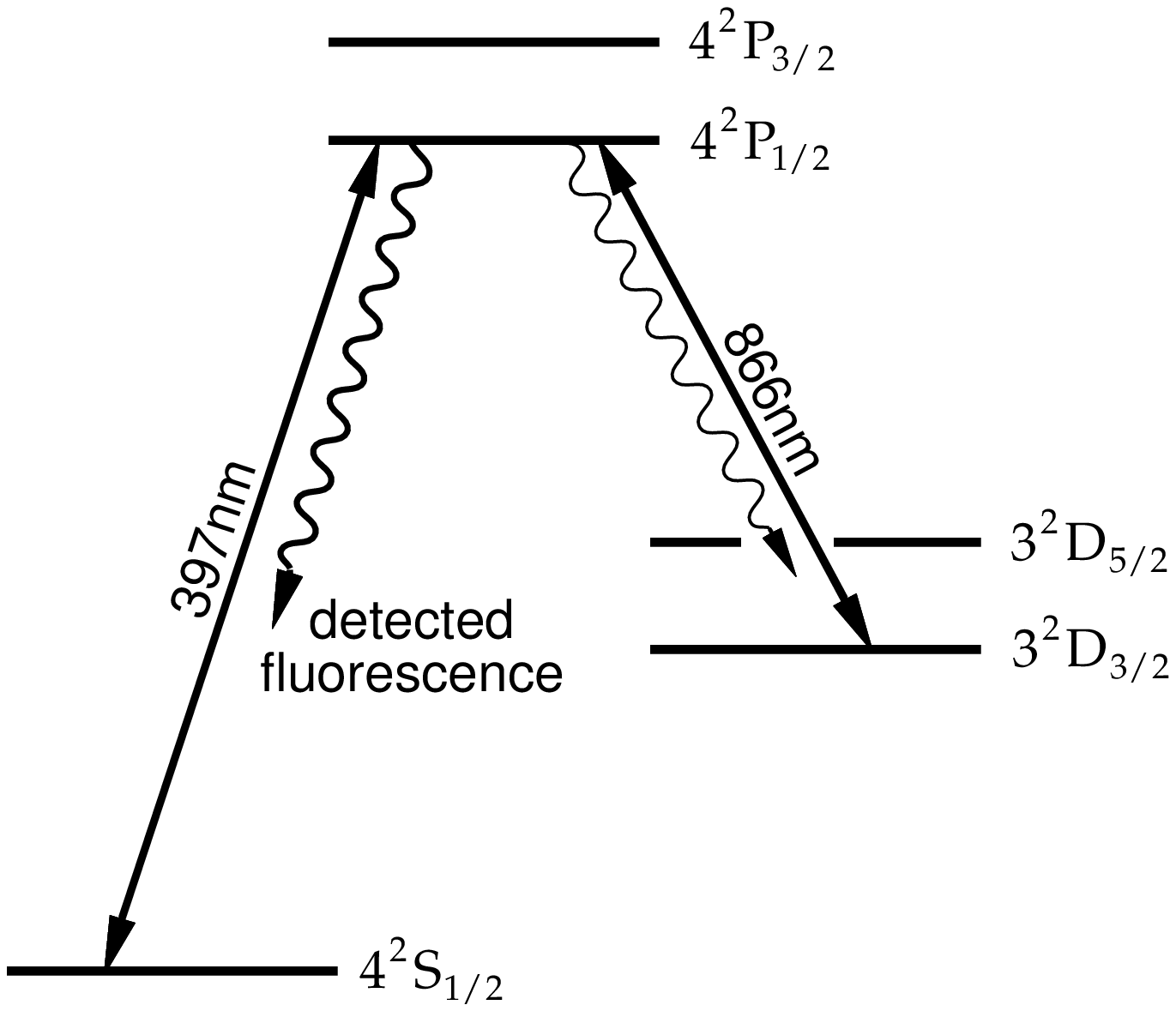}} \noindent
{\bf Figure 3.4(b).} Raman readout, step 2: resonance fluorescence at 397~nm.

\centerline {\includegraphics[width=0.7 \columnwidth]{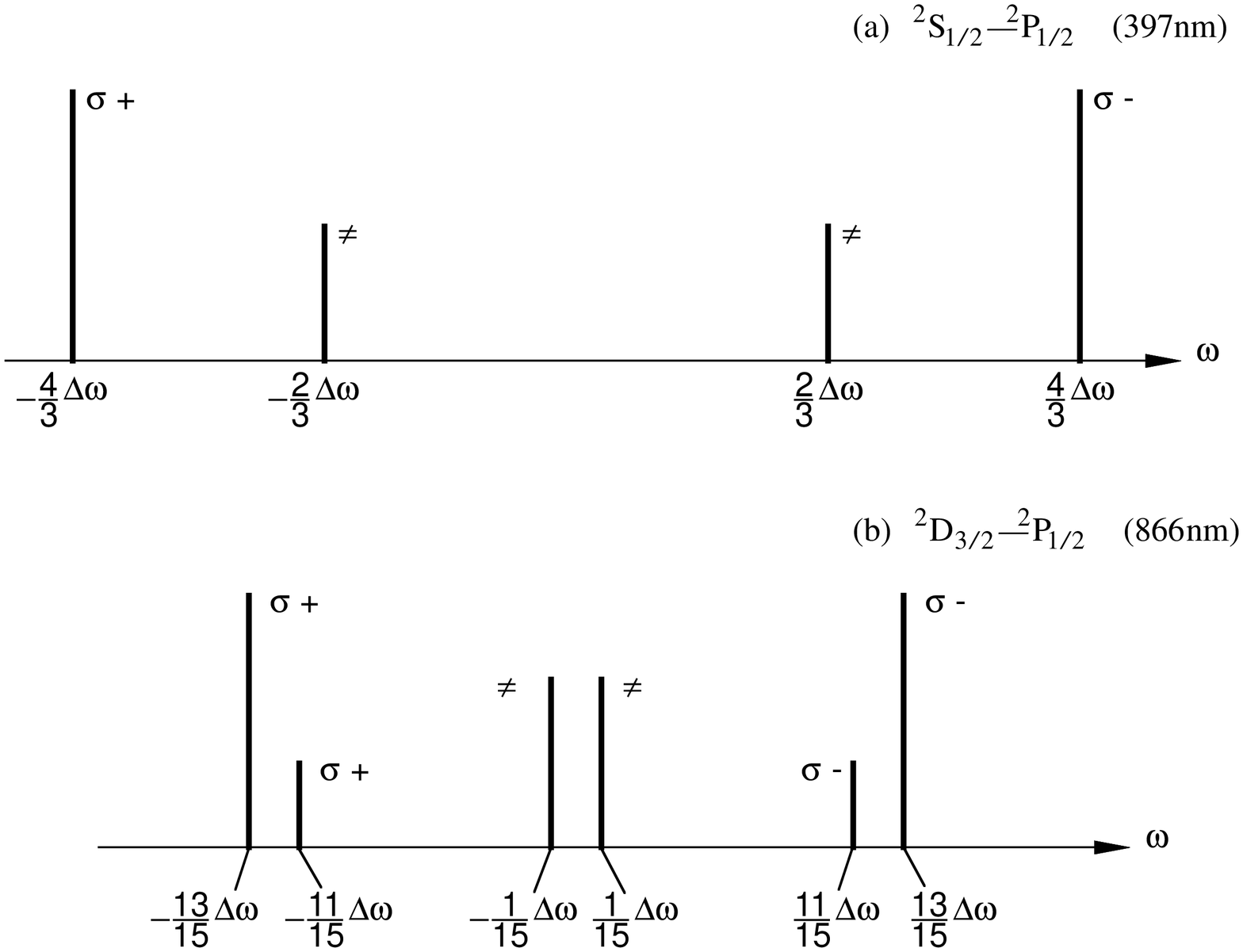}} \noindent
{\bf Figure 4.1.} The Zeeman splitting of (a) the 397~nm $4\,^{2}{\rm S}_{1/2}$ to $4\,^{2}{\rm P}_{1/2}$ transition and (b) of the 866~nm $4\,^{2}{\rm D}_{3/2}$ to $4\,^{2}{\rm P}_{1/2}$. The quantity $\Delta\omega$ is given by the formula $\mu_{B} B/\hbar$, where $\mu_{B}$ is the Bohr magnetron; numerically, $\Delta\omega = (2\pi) 1.4 {\rm [MHz]} B{\rm [Gauss]}$. The resonances represent the following transitions (from left to right, quantum number of lower sublevel stated first):
(a) $m_{J} = +1/2$ to $m_{J}=-1/2$; $+1/2$ to $+1/2$; $-1/2$ to $-1/2$ and  $-1/2$ to $+1/2$; (b) $m_{J} = +3/2$ to $m_{J}=+1/2$ ; $+1/2$ to $-1/2$; $+1/2$ to $+1/2$; $-1/2$ to $-1/2$; $-1/2$ to $+1/2$ and $-3/2$ to $-1/2$.

\centerline {\includegraphics[width=0.7 \columnwidth]{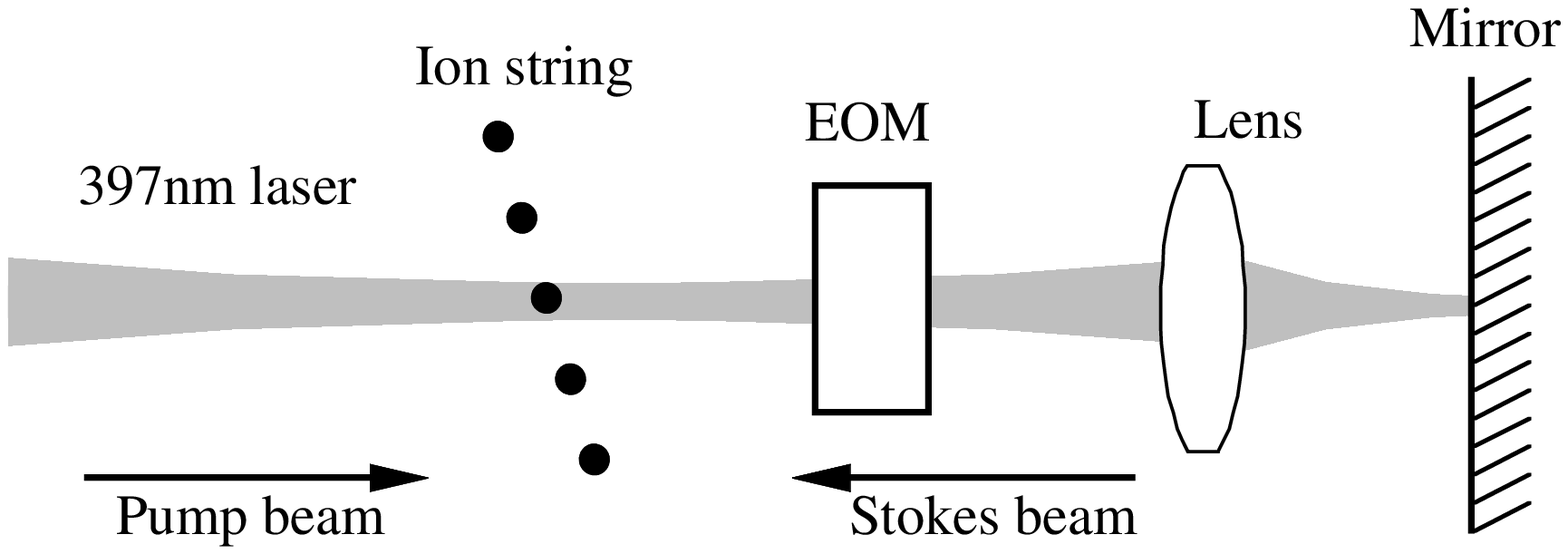}} \noindent
{\bf Figure 4.2.}  Simple experimental layout for producing appropriate Pump and Stokes beams.

\newpage
\section*{Tables}
\noindent
{\bf Table 1.}  Isotopes of Calcium with non-zero nuclear spin and half-lives $\tau$ greater than 1 second.
\begin{center}
\begin{tabular}{|c|c|c|c|}\hline
Isotope&Decay Mode&Half Life, $\tau$&Nuclear Spin, $I$ \\ \hline
$^{41}{\rm Ca}$&Electron Capture&103,000 years&7/2\\
$^{43}{\rm Ca}$&(stable)&-&7/2\\
$^{45}{\rm Ca}$&$\beta^{-}$ emitter&162 days&7/2\\
$^{47}{\rm Ca}$&$\beta^{-}$ emitter&4.5 days&7/2\\
$^{49}{\rm Ca}$&$\beta^{-}$ emitter&8.72 min&3/2\\
$^{51}{\rm Ca}$&$\beta^{-}$ emitter&10 sec&3/2\\
\hline
\end{tabular}\\
\end{center}

\vspace{0.5cm}
\noindent
{\bf Table 2.}  Estimated laser wavelength, power and bandwidth requirements for performing the operations described in this paper. Numbers are intended for general guidance only.  See text for explanation of assumptions and parameters used.
\begin{center}
\begin{tabular}{|c|c|c|}\hline
Laser&Power&Bandwidth\\
\hline
397~nm&0.5~mW& (2$\pi$) 20 MHz\\
866~nm&0.1~$\mu$W&(2$\pi$) 2.0 MHz\\
729~nm&0.2~W&(2$\pi$) 100 kHz\\
\hline
\end{tabular}\\
\end{center}

\end{document}